\documentclass[%
 preprint,
 amsmath,amssymb,
 aps,
]{revtex4-2}

\usepackage{graphicx}
\usepackage{dcolumn}
\usepackage{bm}
\usepackage{enumitem}

\begin{document}


\title{Landmarks and Frontiers\\in Biological Fluid Dynamics}

\author{John O. Dabiri}
\affiliation{%
 School of Engineering, Stanford University\\
 Stanford, California, USA 94305
}%




\date{\today}

\begin{abstract}
Biological systems are influenced by fluid mechanics at nearly all spatiotemporal scales. This broad relevance of fluid mechanics to biology has been increasingly appreciated by engineers and biologists alike, leading to continued expansion of research in the field of biological fluid dynamics. While this growth is exciting, it can present a barrier to researchers seeking a concise introduction to key challenges and opportunities for progress in the field.  Rather than attempt a comprehensive review of the literature, this article highlights a limited selection of classic and recent work. In addition to motivating the study of biological fluid dynamics in general, the goal is to identify both longstanding and emerging conceptual questions that can guide future research. Answers to these fluid mechanics questions can lead to breakthroughs in our ability to predict, diagnose, and correct biological dysfunction, while also inspiring a host of new engineering technologies. 
\end{abstract}

\maketitle


\section{\label{sec:level1}Why Study Biological Fluid Dynamics?}

The field of biological fluid dynamics continues to grow and diversify as researchers discover myriad new ways in which fluid mechanics influences biological systems. Many scientists and engineers are attracted to this area of study solely because of the elegance one often discovers in the underlying flow physics. Yet, even putting aside motivations rooted in the joy of basic science research and the inherent aesthetic of biological flows, the study of biological fluid dynamics has potential for significant, tangible impact. Broadly speaking, the potential impacts can be categorized as (1) effecting the function of naturally-occurring biological systems, and (2) using biological function to inspire new engineering technologies. The context of human health will provide motivation for the former category, while opportunities to address climate change will headline the latter.




\subsection{\label{sec:level2}Human Health: Predicting, Diagnosing, and Correcting Dysfunction}

The top 7 leading causes of death in 2016 were collectively responsible for approximately 45\% of global mortality, accounting for more than 25 million lives lost~\cite{WHO2018}. In each of these 7 classes of mortality~\footnote{Ischaemic heart disease; stroke; chronic obstructive pulmonary disease; lower respiratory infections; Alzheimer disease and other dementias; trachea, bronchus, and lung cancers; and diabetes mellitus}, fluid mechanics plays a fundamental role. Flow-structure interactions in the cardiovascular system are responsible for heart disease and stroke, the two leading causes of death globally for the past 15 years. The long-term presence of particle-laden flow in the lungs\textemdash from air pollution or smoking, for example\textemdash leads to chronic obstructive pulmonary disease (COPD), the third most common cause of mortality. Lower respiratory infections such as pneumonia are the fourth leading cause of death globally and the leading cause of mortality for children in developing countries. In addition to the direct role that fluid mechanics plays in the lungs of those suffering from these respiratory infections, the transmission of the associated communicable diseases involves a complex interplay of multiphase, sometimes non-Newtonian fluids with ambient environmental flows~\cite{Scharfman_etal_2016,Kim_etal_2019}. 

Only in the past several years has it become apparent that other processes associated with significant mortality also have fluid mechanical processes underlying them. The growth of cancerous tumors relies on angiogenesis, the growth of new blood vessels from existing vasculature. That process can be triggered by shear stresses exerted by blood on the surrounding tissues. Furthermore, cancers can spread from an initial, primary site to other locations in the body by advection of cancerous cells in the bloodstream. This process of metastasis, which fundamentally involves the transport and sedimentation of particulates by unsteady flows, is ultimately a fluid mechanical phenomenon~\cite{Koumoutsakos_etal_2013}. 

While the flow of blood through vessels in the brain has long been appreciated for its importance in strokes, the flow of interstitial fluids outside of blood vessels in the brain may be even more important for its function. Recent work suggests that the clearance of waste products via this porous media flow might underlie devastating pathologies such as Alzheimer’s disease~\cite{Bacyinski_etal_2017}. 

Although less common, many other diseases have equally tragic consequences, and they also feature complex flow-structure interactions at a variety of length and time scales. For example, cystic fibrosis affects the ability of cilia located deep within the lungs to transport mucous effectively. As a result, those who suffer from this genetic disorder are more prone to lung infections and also struggle to breathe efficiently. Many digestive disorders are similarly characterized by pathological transport of complex fluids. Given emerging research regarding the impact of the bacteria in the digestive system on brain function~\cite{Mayer_etal_2014}, there are potentially multiple avenues through which fluid mechanics is central to achieving success in the grand challenge of understanding how the brain works.

The prevalence of fluid mechanics in human health suggests that breakthroughs in our understanding of biological fluid dynamics can enable more accurate prediction and earlier diagnoses of dysfunction, both of which are known to improve diseases prognoses. Moreover, a mechanistic understanding of how fluid mechanics mediates biological processes in the human body can inform interventions to correct dysfunction. The impact of success in these efforts is hard to overstate: the cumulative global economic impact of cancer, diabetes, mental illness, heart disease, and respiratory disease is projected to reach 47 trillion USD by 2030 under the status quo~\cite{Bloom_etal_2012}. More importantly, millions of lives can be positively impacted by advances in our understanding of the relevant fluid mechanics.

\subsection{\label{sec:level2b} Using Biology to Inspire New Engineering Technologies}

Since the sketches of Leonardo da Vinci and likely much earlier, humankind has taken inspiration from biological fluid dynamics to invent new technologies. Much of this work was initially focused on directly mimicking the flow-structure interactions observed in nature, as is well documented in the quest for heavier-than-air flight. However, those efforts relying on biomimicry represent a small subset of the methods by which biological fluid dynamics can inspire engineering, and only a sliver of the potential applications have been pursued to date.  

For example, aviation is but one portion of an enormous global transportation sector, which accounts for almost 30\% of global energy consumption at an annual cost approaching 2 trillion USD~\cite{EIA2017}. Given the immense scale of this energy use and its current sourcing primarily from fossil fuels, improvements in the efficiency of transportation are vital to reduce global carbon emissions. Aside from losses in the engine, the energy consumption for transportation is almost entirely associated with fluid dynamic drag. For this reason, significant research efforts are focused on gleaning strategies from animal swimming and flying to reduce drag on engineered vehicles. Empirical evidence suggests that the energy consumed by the transportation sector could be reduced significantly if the performance of engineered systems can approach that of biological locomotion. This can be illustrated by comparing the energy cost of transport (COT; J kg$^{-1}$ m$^{-1}$)\textemdash a common measure of locomotor efficiency\textemdash of various modes found in nature. The most efficient movers in the animal kingdom, jellyfish, exhibit a COT as low as 0.3 J kg$^{-1}$ m$^{-1}$~\cite{Gemmell_etal_2013}. By comparison, the COT of even an efficient automobile like the Toyota Prius is almost four times higher~\cite{Prius2017}. Air transport is another factor of six more expensive than the Prius, approximately 7 J kg$^{-1}$ m$^{-1}$~\cite{Peeters_etal_2005}. In fact, the only known mode of transportation whose COT is lower than jellyfish swimming is via surface ships moving at low speed. The quadratic penalty of aerodynamic drag on COT is avoided by ships when moving at low speed, and wave drag at the free surface is also minimized. The mass in the denominator of the COT further reduces its value for large surface ships (which one could argue is an artifact of the COT not being dimensionless~\cite{Hornung2006}). Nonetheless, the constraint on speed necessarily limits the performance envelope of that mode of transportation. Moreover, the fuel oil used by most surface ships is sufficiently carbon-intensive that international shipping still accounts for more carbon emissions than all but 5 countries~\cite{Olivier_etal_2016}. There are therefore significant benefits to reap in the battle against climate change by reducing fluid dynamic drag on land, in the air, and at sea.

Efforts to reduce global energy consumption are complemented by initiatives to generate energy more sustainably. It is estimated that energy from renewable sources will need to triple by 2050 to limit global temperature rise due to climate change~\cite{IPCC2018}. Innovative methods to generate energy from wind, water motion, and solar irradiance are therefore of paramount importance. Although fluid dynamic energy conversion is not widely observed in nature, the same biological flow-structure interactions that inspire drag reduction can also inform strategies to harvest energy from environmental flows. In the case of wind energy, the performance envelope of the turbine blades can potentially be increased by leveraging knowledge of dynamic stall control observed in flapping flight and lift-based swimming~\cite{Fish_etal_2011}. Moreover, wind turbine siting in large farms could leverage aerodynamic interactions that are similarly exploited in swarms of swimming and flying animals~\cite{Whittlesey_etal_2010}. 

Biological fluid dynamics would be incomplete without acknowledging the rich set of unsteady flow-structure interactions that occur between environmental flows and plant life. Indeed, passive manipulation of aerodynamic loading, a property exhibited by terrestrial and aquatic vegetation~\cite{Vogel1989}, can potentially be exploited for energy harvesting (e.g. for hydrokinetic energy harvesting in currents) or simply to extend the lifetime of energy infrastructure under unsteady fluid dynamic loading (e.g. solar panels in atmospheric turbulence).

As an aside, the ongoing revolution in robotics, powered in part by advances in artificial intelligence, presents new opportunities to leverage biological fluid dynamics in technologies that touch every part of human existence beyond those already discussed. Low-cost, bespoke hardware and software now make it possible to achieve sensing, actuation, and control that rival biological systems. Fluid mechanics researchers are just beginning to use these capabilities to more effectively probe biological systems and to replicate their fluid dynamic functions. It is premature to speculate on the impact of these tools on either of the two categories just discussed, but they suggest that we are entering an era of new possibilities for experimental fluid mechanics in particular. 

\section{\label{sec:level1b}Landmarks and Frontiers}

The preceding section has hopefully convinced the reader of the many merits of a career studying biological fluid mechanics. It remains to provide an orientation to the research landscape in which the reader will be engaging. A comprehensive review is neither feasible nor is it advisable from pedagogical standpoint. Indeed, this written narrative is based on a similarly brief presentation given at the 2018 Annual Meeting of the American Physical Society Division of Fluid Dynamics. It was in preparation for that occasion that the impossibility of summarizing the field of biological fluid dynamics became fully apparent. 

Therefore, in lieu of a comprehensive anthology, the following synthesis represents my own, admittedly biased and incomplete, perspective on the field. The set of papers discussed is intentionally small. I have limited the collection to 10 papers, making it possible to read all of them in a brief but dedicated course of study. The selected literature skews toward work published in the past two decades, which intentionally coincides with an explosion of interest in biological fluid dynamics. These 10 papers are augmented by other references that support the flow of the narrative, but it is important to emphasize that the bibliography is not representative of the breadth or depth of scholarship in the field. Indeed, we regrettably will not discuss classic work from pioneers of the field such as Sir James Gray, Sir James Lighthill, Yuan-Cheng ``Bert'' Fung, and Ted Y. Wu. The reader is encouraged to explore their contributions, as they provide rich scientific and historical context. 

The following narrative is constructed from a taxonomy of biological contexts at multiple length and time scales. The categories are overlapping in some cases, both in terms of biological systems that are featured and fluid mechanics principles that are most pertinent. The discussion will hearken back to the earlier presentation of applications in human health and bio-inspired engineering. Indeed, it is worth emphasizing that each category of biological function below will likely offer something important to both of the broad groups of applications introduced in the previous section. 
\newpage
The categories of biological function that will guide this introduction relate to the role of fluid mechanics in 
\begin{enumerate}[label=\Alph*.]
	\item Locomotion
	\item Sensing
	\item Reproduction
	\item Development
	\item Internal Transport
	\item Biological Communities
\end{enumerate}

With each of these categories I have associated papers that I consider ``landmarks'' in biological fluid dynamics. I use the term in its navigational context, to indicate the value of these papers for orienting the reader to interesting fluid mechanics concepts in that area. To be sure, the field of biological fluid dynamics has a wealth of important literature, so the reader should consider those described below as but a brief glimpse of the exciting work in this field. If this narrative is successful, it will encourage a deeper dive into the broader literature.

The landmarks are each paired with frontiers that hold tremendous promise for breakthroughs in our understanding of fluid mechanics and in our knowledge of biology. 

\subsection{\label{sec:level2a} Locomotion}
Motility is fundamental to life, from cellular to organismal scales. With the possible exception of life forms that persist in solid sediments, all organisms therefore contend with fluid mechanics as they move through their environment. Here we discuss landmarks and frontiers in locomotion and the physical processes by which it is mediated. The primary focus of the present section is on fluid mechanics in the inertial regime; we will return to motility at low Reynolds numbers in our later discussion of biological reproduction.

\subsubsection{The role of wake dynamics}

\noindent\emph{\textbf{Landmark:} ``Flying and swimming animals cruise at a Strouhal number tuned for high power efficiency.'' Taylor, Nudds, and Thomas. Nature 425: 707--711 (2003)~\cite{Taylor_etal_2003}\\
\textbf{Frontier:} Connecting three-dimensional wake vortex kinematics to propulsive efficiency}
\vspace{5mm}

The discussion of costs of transport in Sec. I.B. highlighted the fact that many swimming and flying animals can move significantly more efficiently than current engineered systems. A longstanding question in the study of biological propulsion is how animal appendages manipulate the local flow to achieve that feat. Taylor et al.~\cite{Taylor_etal_2003} identified a potential organizing principle that has guided research into this question over the past decade and a half. They observed that when the animal morphology and kinematics are non-dimensionalized as a Strouhal number, i.e. $St = fA/U$ where $f$ is the frequency of appendage motion, $A$ is its amplitude, and $U$ is the cruising speed, that parameter falls into a relatively narrow range centered between 0.2 and 0.4. This range of Strouhal number coincides with that associated with maximal growth of linear perturbations of a jet~\cite{Triantafyllou_etal_1991}, and therefore it has been speculated that the observation of a narrow range of locomotor Strouhal number reflects an optimization of wake vortex formation.

The conceptual connection from a Strouhal number based on animal kinematics to efficient vortex formation raises several issues that remain unresolved. First, the stability analysis that identifies a Strouhal number for maximum growth of perturbations is linear, whereas vortex formation by animal appendages is a decidedly nonlinear process involving large-amplitude body motions. Second, the stability analysis assumes the existence of a steady base flow. While one can construct such a base flow from the time-averaged wake in order to proceed with the linear stability analysis, strictly speaking, the base flow does not actually exist. Hence, one can question whether the real vortices observed in the wake are truly the product of an amplification of small disturbances. Put more simply, the correspondence between the Strouhal number based on animal kinematics and the Strouhal number from linear stability analysis may be a numerical coincidence.

To be sure, many other arguments have been developed to explain the observation of a narrow range of kinematic Strouhal numbers in animal locomotion, including in very recent literature (see~\cite{Floryan_etal_2018} and references therein). A fundamental issue that requires examination is whether the Strouhal number is indeed the appropriate organizing concept to use in studying efficient locomotion. In this context, the first question that demands a more rigorous conclusion is whether the formation and shedding of vortices is indeed the underlying physical process that dictates the efficiency of locomotion. One can argue that the historical focus on the wake of swimming and flying organisms was in part motivated by an inability to measure the flow directly in contact with the animals. As such, Trefftz plane and control volume analyses, i.e. tracking the flow downstream from the organism, were the only available options for analysis of the fluid dynamics for many decades. The advent of quantitative flow visualization tools like particle image velocimetry enabled measurement of vortex dynamics~\cite{Lentink_etal_2009}, which brought us a step closer to the flow-structure interactions that create locomotive forces~\cite{Muijres_etal_2008,Hubel_etal_2016}. However, analyses of wake vortices are also an indirect evaluation.

A direct, causal relationship between wake vortex formation and efficient locomotion has been sought in research focused on the formation of individual vortex rings. Gharib et al.~\cite{Gharib_etal_1998} introduced the concept of the vortex formation number, a non-dimensional timescale at which individual vortex rings attain maximum circulation. Subsequent measurements~\cite{Krueger_Gharib_2003} demonstrated that the thrust generated by a starting jet is created most efficiently when the duration of each jet pulse matches the vortex formation number. 

The aforementioned results regarding vortex formation number have not been adopted widely in the study of animal swimming and flying, perhaps because of uncertainty regarding whether the wakes generated during biological propulsion can be adequately modeled as vortex rings governed by the formation number concept. Nonetheless, it is worth further investigating the observation that not only is the inverse of the vortex formation number a parameter that physically parallels the kinematic Strouhal number, but its quantitative value of 4 is consistent with the range of kinematic Strouhal numbers $St = 0.2-0.4$ identified by Taylor et al.~\cite{Taylor_etal_2003} in swimming and flying animals. 

A logical next step in this analysis would be construction of a compilation of the vortex formation numbers for a variety of swimming and flying organisms, similar to that created using the Strouhal number. The challenge in doing this is that, unlike the kinematic Strouhal number that can be estimated based on observations of the animal alone, the formation number requires simultaneous measurement of the animal motion and its three-dimensional flow field. As measurement capabilities in experimental fluid mechanics become more advanced, the pursuit of time-resolved, three-dimensional measurements of animal-fluid interactions has become a promising frontier in the study of efficient animal locomotion.

Empirical measurements should be complemented by new analyses that extend our understanding of vortex formation beyond the simplified geometries of vortex lines and rings, to accommodate the full complexity of real animal wakes. In the limit of inviscid vortex lines, the vortex dynamics can be computed via Biot-Savart induction~\cite{Saffman1992}. If the concept of vortex formation number can be described within this framework of inviscid vortex dynamics, then it might also be possible to apply that knowledge to the dynamics of vortex wakes at finite Reynolds number. Although vortex dynamics at finite Reynolds numbers do not admit such straightforward approaches as Biot-Savart induction, the observed robustness of formation number to Reynolds number~\cite{Rosenfeld_etal_1998} suggests that an inviscid approximation of animal wakes may be sufficient to understand the connection with efficient locomotion~\cite{OFarrellDabiri2014}. 

One particularly promising avenue in this regard could be application of the Kelvin-Benjamin variational principle, which describes energetic conditions satisfied by vortex rings at the formation number~\cite{Kelvin1880,Benjamin1976}. While the Kelvin-Benjamin variational principle has primarily been applied to constrain the evolution of vortex rings with finite-size cores~\cite{Dabiri2009}, extension to infinitely thin vortex loops might provide the means to apply the vortex formation number concept to arbitrary animal wakes.

\subsubsection{The role of body dynamics}

\noindent\emph{\textbf{Landmark:} ``Interactions between internal forces, body stiffness, and fluid environment in a neuromechanical model of lamprey swimming.'' Tytell, Hsu, Williams, Cohen, and Fauci. Proc. Natl. Acad. Sci. 107: 19832--19837 (2010)~\cite{Tytell_etal_2010}\\
\textbf{Frontier:} Elucidating the roles of active and passive body stiffness in propulsive efficiency}
\vspace{5mm}

Implicit in the preceding section is a long-held assumption that the key to efficient locomotion can be revealed by studying the vortex wake produced by a swimming or flying animal. This view may be incorrect, or at least incomplete. Specifically, while the wake will reflect the net force that an organism exerts on the surrounding fluid, the wake does not necessarily indicate the total energetic cost of locomotion for the animal. The latter depends on the power exerted by the animal musculature, which in turn depends on the body's own resistance to deformation. Some of that resistance will be elastic, e.g. storing energy in bending motions that can be recovered later in the propulsive cycle of the appendages~\cite{Eldredge_etal_2010}. Another portion will be inelastic and ultimately lost to heat~\cite{Biewener2003}. 

The cost of transport parameter discussed previously reflects a combination of energy losses due to actuation (i.e. the inelastic resistance to body deformation) as well as kinetic energy lost to the animal's wake. Therefore, animals that we celebrate for outperforming our engineered vehicles in terms of cost of transport might only be exhibiting superior actuation as opposed to superior fluid mechanics. If so, then attempts to engineer more efficient propulsion systems by mimicking the fluid mechanics of animal locomotion might be futile.

The essential role of body dynamics in efficient locomotion was highlighted by Tytell et al.~\cite{Tytell_etal_2010}, who used numerical simulations to artificially change the stiffness of animals swimming \emph{in silico}.  They observed a relationship between body stiffness and propulsive efficiency that likely matches your intuition: if the body is too pliant, significant energy is lost to deformation of the body instead of being used to affect the momentum of the surrounding fluid. In contrast, an overly stiff body cannot undergo the kinematics necessary to generate sufficient fluid motion. An intermediate level of body stiffness strikes an efficient balance, leading to optimal propulsive efficiency. Underscoring the preceding discussion, Tytell et al.~\cite{Tytell_etal_2010} observe that the same swimming kinematics (and hence the same wake dynamics, assuming identical initial flow conditions) can be achieved by multiple combinations of muscle force and body stiffness. Hence the wake associated with a given set of swimming kinematics is not uniquely connected to the energetics of locomotion. We can only draw conclusions about the total energetic efficiency of locomotion if the internal mechanical losses are small relative to fluid dynamic losses. This conclusion suggests that a frontier in the study of animal locomotion is to deduce how the energy budget for locomotion is partitioned between actuation and wake losses. Advances in combined respirometry and flow velocimetry can provide one route for progress. Extending coupled flow-structure interaction solvers toward three-dimensional, muscle-actuated simulations represents another promising avenue, as does the use of mechanical analogues~\cite{Leftwich_etal_2012}.

An open question is whether there exists a parameter space that explicitly incorporates both the swimming kinematics of Taylor et al.~\cite{Taylor_etal_2003} and the body mechanics of Tytell et al.~\cite{Tytell_etal_2010}. The latter work demonstrated that the total energetic efficiency of locomotion is not uniquely determined by the Strouhal number; indeed they find that similar Strouhal numbers can correspond to power coefficients that vary by a factor of three. This lends further credence to the possibility that the observed Strouhal convergence is either a numerical coincidence, or that it represents an amalgam of even more fundamental physical phenomena, some\textemdash but not all\textemdash of which are captured by the kinematic Strouhal number. 

\subsubsection{The frontier of maneuverability}

Despite the wide diversity of animal systems that have been studied to understand the fluid mechanics of swimming and flying, one common thread has been a primary focus on unidirectional locomotion. Perhaps because experiments in wind tunnels and flumes generally confine flying and swimming to a single direction, and numerical models are most convenient in a similar scenario, much of what we know about the fluid mechanics of animal locomotion is limited to this subset of behaviors. In contrast, ecological studies of locomotion have long appreciated that animals are constantly changing direction, whether to catch prey, avoid predators, or find mates for reproduction~\cite{Seuront_etal_2004}.

The fluid mechanics of turning have received much less attention, despite their importance for understanding natural locomotion and for designing engineering vehicles that can achieve a similar performance envelope. Hence, this topic still awaits a ``Landmark'' paper. In very recent work, we examined subtle body kinematics that allow inertial swimmers to maximize their moment of inertia during torque generation, and then to subsequently minimize their moment of inertial for more rapid turning~\cite{Dabiri_etal_2019}. This strategy was observed in two very different lineages of animals, jellyfish and zebrafish, which hints at the possibility that this approach can be generalized to many other body forms.

Regardless of the generality of the preceding results, an important frontier in the study of animal locomotion will be to embark on comprehensive studies of maneuverability to complement the rich literature on unidirectional locomotion. Whereas the latter studies have been able to exploit one- and two-dimensional simplifications, the study of six degree-of-freedom maneuverability will typically require that we capture the full three-dimensional dynamics of the animals and the surrounding fluid. As mentioned above, experimental fluid mechanics is just beginning to approach this capability; numerical simulations will likely achieve results sooner~\cite{Ren_etal_2016}, and can therefore provide guidance for experiments. Is it possible to achieve a grand unifying theory of animal locomotion, one that explains the tight constraint on kinematic Strouhal number, incorporates the role of body actuation, and that can be generalized to arbitrary three-dimensional trajectories? One promising direction is to leverage fluid mechanics in the inviscid limit, where the fluid motion and body kinematics are tightly coupled via added-mass dynamics~\cite{Eldredge2019}. Such methods admit three-dimensional models of vortex dynamics with less computational expense than is associated with direct numerical simulations. While these approaches may be less accurate, they can potentially capture qualitative trends such as the existence (or absence) of locomotor kinematics that optimize efficiency, maneuverability, stealth, or other commonly sought performance characteristics in bio-inspired propulsion.

\subsection{\label{sec:level2b} Sensing}

The ability of an organism to move is often mediated and motivated by sensory inputs from the environment. Those sensing capabilities are themselves either facilitated or limited by the surrounding fluid. An improved understanding of the fluid dynamics of sensing can therefore enable a more complete understanding of biological systems, while also inspiring more capable sensing technologies such as biochemical environmental monitoring. It is beyond the scope of this paper to discuss the full variety of sensing modalities that arise in nature. In particular, there is a large and growing literature on hearing (i.e. acoustic sensing) which will not be discussed here~\cite{Mittal_etal_2013,Obrist2019}. However, below we examine two broad classes that each present exciting opportunities for science and engineering.

\subsubsection{Chemical sensing}

\noindent\emph{\textbf{Landmark:}  ``Lobster sniffing: antennule design and hydrodynamic filtering of information in an odor plume.'' Koehl, Koseff, Crimaldi, McCay, Cooper, Wiley, and Moore. Science 294: 1948--1951 (2001)~\cite{Koehl_etal_2001}\\
\textbf{Frontier:} Deducing three-dimensional scalar field evolution from sensors with limited spatiotemporal bandwidth}
\vspace{5mm}

Predators, prey, and potential mates each release chemical cues into the surrounding fluid, either passively or by active means. In the absence of complex environmental flows, the fate of those cues would be relatively straightforward to predict, and tracking a cue to its source would be trivial. Reality is not so simple, as the combination of locomotion by the source of the chemical cue and unsteady environmental flows together make the task of chemical sensing quite challenging. Compounding matters is the relatively limited bandwidth of the biological sensors, both spatially (e.g. due to sensors being limited to a few locations on the organism body) and temporally (e.g. due to limited sampling frequency and memory effects from previous exposure to chemical signals). 

Koehl et al.~\cite{Koehl_etal_2001} identified another fundamental constraint on chemical sensing: the formation of a viscous boundary layer of fluid around the chemical sensors themselves. When the Reynolds number corresponding to the sensor motion is low\textemdash a likelihood given the small size of the hairlike sensors\textemdash a large region of fluid persists in the vicinity of the sensor as it moves. Therefore, rather than detecting the changing chemical concentrations in the environment as an animal moves through it, the animal instead detects the concentrations in the same fluid trapped near the sensor surface. 

To be sure, the chemical concentrations in the fluid trapped in the boundary layer can change via molecular diffusion. The limitations of this process can be appreciated by comparing the diffusion of chemical cues with the diffusion of momentum into the viscous boundary layer. This ratio of viscous diffusion coefficient $\nu$ to chemical diffusion coefficient $D_{c}$, given by the Schmidt number $Sc = \nu/D_{c}$, is of order 1 in air. Hence, the air in the vicinity of a moving organism will be set into motion by viscous forces at approximately the same time that any chemical cues from the organism arrive at that location in the fluid (assuming that a gradient in chemical cues exists normal to the animal body). Conversely, just as the animal begins to detect the chemical concentrations in the fluid environment in which it has arrived, the viscous forces that it exerts on the flow will cause that portion of the surrounding fluid to move, thereby potentially changing its chemical concentrations. This situation is exacerbated by orders of magnitude in water, which typically exhibits Schmidt numbers of order 10$^{3}$. In this case, fluid mechanics dominated by viscous diffusion will cause the environment to be perturbed by the motion of the physical sensors long before the sensors have acquired information about chemical composition of the environment.

A clever solution to this conundrum was identified by Koehl et al.~\cite{Koehl_etal_2001} in their study of lobster antennae. Specifically, the sensor kinematics can be adjusted such that advection dominates the transport of both momentum and scalars (e.g. dissolved organic matter) in the surrounding fluid instead of diffusion. Quantitatively, rapid motion $U$ of the antenna with characteristic size $L$ corresponds to higher Reynolds number ($Re = UL/\nu$) and Peclet number ($Pe = UL/D_{c}$) regimes, where advection dominates over diffusive transport of momentum and chemical signals, respectively. The asymmetric, time-varying flow-structure interaction of the sensor with surrounding fluid effectively flushes the sensor, so that it can detect fresh signals from its new environment as it moves.

One can anticipate that very close to the chemical receptors, the no-slip condition will cause the relative velocity, and hence the Reynolds and Peclet numbers, to remain small. At this scale, it may be possible that the organism intentionally misaligns gradients in velocity and chemical composition, such that the viscous disturbance of the flow is in a direction that does not affect the chemical gradient being sensed. It may also be the case that at this small scale, advective perturbations to the background chemical gradient (e.g. due to the pressure field of the accelerating antenna) are more significant than viscous perturbations. How organisms account for this advective disruption of the chemical field being sensed remains largely unresolved and ripe for study~\cite{Mellon2007}.

An exciting frontier in the fluid mechanics of chemical sensing is to map the three-dimensional, temporal evolution of ecologically-relevant scalar fields as those environments are interrogated by swimming and flying animals~\cite{Page_etal_2011a,Page_etal_2011b}. To what extent does animal locomotion perturb the signal being analyzed, and what other strategies besides those elucidated by Koehl et al.~\cite{Koehl_etal_2001} are at work to enhance sensing capabilities? Moreover, how does background environmental flow limit the spatiotemporal resolution of sensing that can be accomplished? How is the information gathered from fluid dynamic sensing used for navigation? And finally, in predator-prey interactions or in the competition for mates~\cite{Moore_etal_1999}, how can chemical cues be enhanced, obfuscated, or spoofed for evolutionary advantage?

Webster and Weissburg~\cite{WebsterWeissburg2009} provide an excellent roadmap for research in the context of aquatic organisms, and powerful new laboratory tools continue to emerge, especially at the microscale~\cite{Ahmed_etal_2010,Pekkan_etal_2016,Xu_etal_2018}. The realm of bacterial sensing in particular has been recently shown to exhibit remarkable capabilities. These include chemical gradient sensing at the limit of molecule counting noise~\cite{Brumley_etal_2019} and force-independent measurement of local flow~\cite{Sanfilippo_etal_2019}. This latter ability to measure local fluid strain rate independent of local fluid viscosity could inspire similarly capable engineered sensors that can function in complex fluids with multiple components or even multiple phases. To do so will require a better understanding of how bacteria accomplish that feat. 

As is increasingly the case in biological fluid dynamics, numerical simulations are outpacing our ability to make in situ measurements of these chemical sensing phenomena, a predicament that must be resolved in order to ensure that the predictions from computational models faithfully represent the physics and biology as they occur in nature. At the same time, the behavioral responses of organisms to the sensed environment can be more challenging to model, and empirical observations can lead the way in providing new insights.

\subsubsection{Flow sensing}

\noindent\emph{\textbf{Landmark:}  ``Biomimetic survival hydrodynamics and flow sensing.'' Triantafyllou, Weymouth, and Miao. Annual Review of Fluid Mechanics 48: 1--24 (2016)~\cite{Triantafyllou_etal_2016}\\
\textbf{Frontier:} Using fluid dynamics for object recognition}
\vspace{5mm}

Not only can organisms detect the chemical cues left by living and inanimate objects in the fluid environment, they can also detect their fluid dynamic signatures. As mentioned in the previous section, viscous forces will transfer motion of solid objects to the surrounding fluid, with that information regarding the presence of the object propagating at a speed $U_{\nu}$ set by the kinematic viscosity $\nu$ of the fluid. The characteristic propagation speed decreases with increasing time $t$ and distance $\delta$ from the object, i.e. $U_{\nu} \sim \sqrt{\nu/t}$ or $U_{\nu} \sim \nu/\delta$. For example, once viscosity has transferred the motion of a body to a distance one millimeter away in water, subsequent propagation into the fluid occurs at a speed of approximately a millimeter per second. When the information has traveled a centimeter into the fluid, the rate of propagation of the velocity disturbance will have decreased by an order of magnitude. 

In contrast, the pressure field created by the object will travel at the speed of sound, approximately 1500 m s$^{-1}$ in seawater. Hence, for the regime of flows of relevance to biology, flow sensing will be facilitated primarily by fluid motion induced by the object's pressure field. This presents the opportunity to use a powerful suite of tools from complex analysis, which are most appropriate in the limit of inviscid, potential flows. The methods enable solution of the idealized flow around relatively complex objects based on a set of analytical solutions for simpler geometries. For flows that can be represented in two dimensions, conformal mapping can be used to accomplish this task. Triantafyllou et al.~\cite{Triantafyllou_etal_2016} present an excellent summary of these approaches. Notably, information regarding the size and orientation of an object in two-dimensional potential flow decays with the square of the distance from a pressure sensor. Information regarding the object's position decays more rapidly, as the cube of the distance from the sensor. More specifically, the sensor in this model flow must be within 6 body diameters of the object to detect its size, and within 3 diameters to detect its position. Information regarding the object shape can be deduced by traveling around the perimeter of the target.

While the model ansatz of a potential flow enables the aforementioned quantitative predictions, we must ultimately return to the question of how these processes occur in nature. In some fish, the flow sensors have been well characterized. A set of hairlike sensors are divided into a group embedded in canals along lateral lines on the body surface, appropriately called canal neuromasts; and sensors that protrude from the skin into the flow, called superficial neuromasts. The canal neuromasts are believed to sense pressure, whereas the superficial neuromasts can be deflected by the flow and thereby transduce flow speed into neural signals.

How these signals are used to influence animal behavior is an area of active research. Recent studies in zebrafish suggest the fascinating possibility that differential sensing along the lateral line perimeter could be used to deduce the local flow vorticity via application of Stokes' theorem~\cite{Oteiza_etal_2017}. In this scenario, the loop integral of the local flow, sensed by the superficial neuromasts located on the perimeter of the fish, would be used to determine the tendency of the local fluid to rotate. By swimming toward regions of the low vorticity, animals were observed to maintain a more stable orientation.

The generality of this strategy is not yet known, nor it is clear how it would be affected by the presence of highly unsteady or turbulent flow. A promising frontier in flow sensing is to deduce strategies that can use sparse sensor data from the surrounding flow to inform navigation~\cite{Krieg_etal_2019} and possibly even three-dimensional reconstruction of the environment. A major challenge in this effort is that the potential flow models described earlier cannot be directly used in the case of finite-Reynolds number flows, where viscous effects and flow separation can be common. If first-principles approaches are insufficient, it may be possible to leverage recent advances in machine learning to automate flow sensing based on a library of known correlations between objects and their fluid dynamics signatures~\cite{Mohren_etal_2018, Colvert_etal_2018}.

Once these fluid mechanics are resolved, the ongoing revolution in rapid prototyping, additive manufacturing, and novel materials can be leveraged to create engineered fluid dynamics sensors that are even more capable then their natural analogs.

\subsection{\label{sec:level2c} Reproduction}

\noindent\emph{\textbf{Landmark:}  ``Life at low Reynolds number.'' Purcell. American Journal of Physics 45: 3--11 (1977)~\cite{Purcell1977}\\
\textbf{Frontier:} Identifying strategies to avoid or exploit time-reversibility at low Reynolds numbers}
\vspace{5mm}

Life begins at low Reynolds numbers. For the majority of non-plant biomass on earth, represented by bacteria, phytoplankton, and other microorganisms, life also persists at low Reynolds numbers. Larger organisms eventually transition to our inertial world, but the physical mechanisms that facilitate their sexual reproduction are also dependent on the ability of cells to navigate in an environment dominated by viscous forces. For example, for a sperm cell to fertilize an egg in humans, it often must first travel on the order of 4000 body lengths through the cervix, across the uterus, and to the fallopian tubes. It is estimated that up to two-thirds human infertility cases related to conception are due at least in part to failure of sperm cells to make this trek successfully~\cite{Mayo2018}.

Cilia in the fallopian tubes are thought to play a role in the transport of egg cells both before and after fertilization. Although the relative contributions of ciliary action and muscle contraction have not yet been been resolved, low Reynolds number fluid mechanics likely plays a key role in both mechanisms. Dysfunctions in this low Reynolds number transport accounts for up to two-thirds of female infertility in some countries~\cite{Audu_etal_2009}.

An improved understanding of the physical mechanics underlying the reproductive process can enable breakthroughs in the diagnosis and treatment of infertility. However, the pertinent fluid mechanics can be somewhat unintuitive. Recognizing the disconnect between life at low Reynolds numbers and our daily experience in an inertial world, Purcell~\cite{Purcell1977} provides a delightful presentation of the unique fluid mechanics that govern processes at the microscale.

The consequences of a world without inertia are perhaps most apparent in the context of locomotion. In our earlier discussion of that topic, we explored the notion that motion of fluid in the wake of an organism could be used to infer the physical forces that the organism had previously exerted on the fluid during locomotion. Indeed, the presence of inertia allows fluid motion to persist beyond the moment of its initial creation, such that we can observe its effect in the form of wake vortices. Another, less appreciated consequence of the fluid inertia is that the larger the magnitude of the force on the fluid (e.g. by accelerating the appendage more forcefully), the longer that the resulting fluid motion can persist. This enables nonlinear interactions between the flows created by the same appendage at successive instants in time. Even simple locomotor kinematics can create complex flows due to the inherent time-dependence associated with inertial flows~\cite{LipinksiMohseni2009}. 

When the Reynolds number is sufficiently low, those inertial processes vanish. As soon as the body motion ceases, so too does the fluid motion. Under these circumstances, the instantaneous motion of the fluid is determined by the instantaneous motion of the body, regardless of the time-history that led to the current state. Given the time-independence of the flow, we can more effectively describe locomotion in the limit of vanishing Reynolds numbers as a sequence of time-independent body shape configurations. The fluid displacement that results from a given sequence of body shape configurations can be computed by integrating the flow that occurs during each body shape over the sequence of distinct body shapes. Importantly, the sequence-integrated flow will be equal in magnitude but opposite in direction if the same sequence is executed in reverse order.

Reciprocal motions, which by definition comprise a motion sequence that is immediately followed by its execution in reverse order, are therefore incapable of creating net locomotion. Instead, net locomotion at low Reynolds numbers requires non-reciprocal kinematics. Two examples commonly found in nature are the corkscrew, which returns to its initial configuration without a reversal of the motion sequence; and asymmetric paddling, for which the shape of the propulsive appendage is different during the initial motion and the return stroke. The former is exemplified by some microorganisms that use flagella for propulsion. The latter is commonly found in cilia used for locomotion or for fluid transport.

The literature on low Reynolds number processes in biology is simply too voluminous to do full justice here. However, even this brief introduction to the constraint of time-reversibility in the context of reproduction helps us to appreciate key research frontiers in biological fluid dynamics at the microscale. For example, a deeper dive into the governing equations reveals that application of the low Reynolds number assumption to real biological systems is not always a straightforward task. Consider, for example, the oscillatory waving of an appendage of length $L$ at amplitude $A$ and wavelength $W$, where $A \ll L \ll W$. The characteristic Reynolds numbers based on $A$, $L$, and $W$ may not all satisfy the requirement of being vanishingly small. In such cases, it might be appropriate to retain the time-derivative of the velocity field and possibly also the nonlinear advection term in the Navier-Stokes equations, lest we lose essential physics. A classic example of this scenario is pulsatile flow in small blood vessels or airways. The flow physics in these cases is governed not only by the Reynolds number (e.g. based on the mean flow speed) but also by the Womersley number $Wo = L(\omega / \nu)^{1/2}$, where $L$ is often taken as the tube radius and $\omega$ is the frequency of the pulsatile flow~\cite{Womersley1955}.   

In some cases, direct observations of the associated flow field can confirm the absence of inertial effects. However, in practice such an approach can be misleading. This is because there exist at least two biologically-relevant mechanisms that can lead to net locomotion from reciprocal motions at low Reynolds numbers, in apparent contradiction to the physical arguments above.  First, locomotion within a bounded domain, particularly where the domain boundaries impart their own motion to the fluid, can both enhance non-reciprocal locomotion and also facilitate net propulsion by reciprocal locomotor kinematics~\cite{SpagnolieLauga_2012,LiArdekani2014,Lauga2016}. These boundary effects are relevant in reproduction, as spermatazoa will encounter the boundaries of reproductive organs during the trek toward the target ovum. Transport of ova in the fallopian tubes is significantly influenced by the presence of the nearby tube walls. As noted earlier, these fluid dynamics affect reproductive success. Second, the nonlinear rheology of reproductive fluids can embed memory of previous flow-structure interactions, thereby breaking the time-reversal symmetry that occurs for Newtonian fluids at low Reynolds numbers~\cite{Qiu_etal_2014}. 

Recalling our previous discussion of internal forces during locomotion (Sec. II.A.2), it is also possible that the forces responsible for actuation and the structural response of the body can affect the resulting propulsion in nontrivial ways~\cite{Yu_etal_2006,Sznitman_etal_2010}. Indeed, some classes of infertility are linked to dysfunction of the motor units that drive motion of the sperm flagellum or of the cilia lining the reproductive tract. Scientists have recently developed the ability to artificially create spermatids, immature version of sperm cells that cannot swim~\cite{Zhou_etal_2016}. The absence of locomotion limits their application to in vitro fertilization. However, a rigorous understanding of the aforementioned flow-structure interactions could inform the design of a fully swimming artificial sperm cell, which would significantly broaden the options for treatment of male infertility.

Beyond applications in reproduction, the same technological advances in microelectromechanical systems (MEMS) that promise to enable the biomimetic sensing capabilities discussed in the previous section can also one day enable a generation of engineered, low Reynolds number micromachines~\cite{Dreyfus_etal_2005}. By taking a cue from natural systems, which have adapted to the rules of life at low Reynolds numbers, bio-inspired technologies can similarly exploit these unique physical rules for a variety of biomedical interventions and other applications.

\subsection{\label{sec:level2d} Development}

\noindent\emph{\textbf{Landmark:}  ``Intracardiac fluid forces are an essential epigenetic factor for embryonic cardiogenesis.'' Hove, Koster, Forouhar, Acevedo, Fraser, and Gharib. Nature 421: 172--177 (2003)~\cite{Hove_etal_2003}\\
\textbf{Frontier:} Elucidating the role of fluid mechanics in mechanotransduction}
\vspace{5mm}

The story of reproduction is incomplete if the new life forms do not develop properly. An area of significant and ever-expanding research activity is the study of how environmental factors aid in translation of the genetic code of a fertilized egg into a fully-formed adult. The role of biochemical processes still dominates the attention of researchers in this field, but it has been increasingly appreciated that physical factors play an essential role as well. A key challenge in understanding the role of mechanical forces in development is the seemingly significant difference in relevant scales. Analysis of fluid dynamics relies on the success of a continuum approximation that obviates the need to resolve intermolecular forces. In contrast, gene expression is by definition a molecular process. Early theories of biological development suggested that biochemical factors are sufficient to guide the process. Spatial gradients in biochemical factors alone could trigger local pattern formation, and ultimately, the three-dimensional shapes of organisms~\cite{WolpertTickle2010}.

To demonstrate an essential role for fluid mechanics in development, Hove et al.~\cite{Hove_etal_2003} studied the development of the zebrafish heart. Beginning as a single tubular conduit, the heart subsequently forms a chambered structure that resembles a simplified version of our own cardiac system. Although flow through the heart is essential in adult forms, as it transports oxygen and nutrients to the other organs, the embryonic zebrafish is sufficiently small that it can be sustained by diffusion alone. Hence, Hove et al.~\cite{Hove_etal_2003} could block the flow through the tubular heart and the animal remained alive and continued to grow. However, in the absence of blood flow through the heart, that structure failed to develop chambers, valves, or the general morphology of an adult heart.

These elegant experiments demonstrated that blood flow plays a physical role in development of the heart, independent of the function of blood flow as a transport mechanism. Interestingly, both pressure and shear were observed to play distinct roles, not unlike the canal and superficial neuromasts that we discussed in the context of flow sensing. To be sure, the role of shear had been observed using in vitro studies, which showed that cells at the tissue surface will tend to align with the direction of flow~\cite{Fung1993}. Here, Hove et al.~\cite{Hove_etal_2003} observed that shear stresses two orders of magnitude larger than in previous studies can occur \emph{in vivo}, due to the unsteady nature of the flow. These mechanical signals have sufficient range to trigger a variety of different stretch- and shear-sensitive receptors, which are thought to initiate a cascade of biochemical signals that could create spatial biochemical gradients and associated pattern formation.

Much remains to be discovered regarding the process of mechanotransduction in biological systems and the myriad ways in which it can be manifested~\cite{Sanfilippo_etal_2019}. A frontier in this research is development of new experimental tools that can enable measurement of the fluid flow, calculation of the associated pressure and shear forces, and detection of changes in gene expression that result from the fluid mechanics. 

Challenges abound. Observation of the fluid flow was possible in the work of Hove et al.~\cite{Hove_etal_2003} only because embroyic zebrafish are essentially transparent. Hence, the flow could be visualized using particle image velocimetry and confocal microscopy. Most organisms do not provide this convenience, making it more challenging to measure the pertinent fluid dynamics. We will return to this limitation in the next section on internal flow.

Even where the flow kinematics can be measured, translating the fluid motion into dynamical quantities such as pressure and shear is non-trivial. In principle, the pressure field can be inferred from the velocity field by direct integration of the Navier-Stokes equations. In practice, however, measurement errors accumulate along the integration paths, leading to large error in the resulting pressure calculation~\cite{LiuKatz2006,vanOudheusden2013}. Alternatively, the Navier-Stokes equations can be reformulated to extract a Poisson equation for the pressure. However, solution of the equations requires knowledge of the boundary conditions, and they are rarely known a priori in viscous flows.

Finally, one would ideally like to correlate the fluid dynamic forces with gene expression. Single-cell RNA sequencing now makes it possible to identify signatures of gene expression at the level of individual cells, however this process requires disaggregation of the tissue of interest~\cite{Briggs_etal_2018}. To connect the gene expression in a cell with the fluid dynamics forces it experienced, it will be necessary to reference each cell to its original position in the intact tissue. How to do so reliably remains an open question. Moreover, when cells are exposed to unsteady flow, the timescale of the fluid dynamic forces might be shorter than the response time for gene expression. In these cases, it will be more challenging to infer a direct connection between a specific fluid dynamic regime and the corresponding gene expression.

In spite of these difficulties, the benefits of success would be enormous for our understanding of how fluid mechanics facilitates development. That knowledge could enable earlier diagnosis of dysfunctions related to fluid flow, and possibly even new interventions to correct developmental deformities before they become permanent.

\newpage

\subsection{\label{sec:level2e} Internal Transport}

\noindent\emph{\textbf{Landmarks:}  ``The paravascular pathway for brain waste clearance: current understanding, significance and controversy.'' Bacyinski, Xu, Wang, and Hu. Frontiers in Neuroanatomy 11: 101 (2017)~\cite{Bacyinski_etal_2017}\\
\indent\indent\indent\indent\indent ``Unidirectional pulmonary airflow in vertebrates: a review of structure, function, and evolution.'' Cieri and Farmer. Journal of Comparative Physiology B 186: 541--552 (2016)~\cite{CieriFarmer2016}
\\
\textbf{Frontier:} Discovering new roles for fluid transport in human physiology}
\vspace{5mm}

Some of the earliest scientific studies of biological fluid dynamics occurred in the context of internal flow in the human body. Daniel Bernoulli, whose namesake principle relates velocity to pressure in idealized flows, is surmised to have been inspired by his early training in medicine~\cite{Quinney1997}. The research literature on blood flow in the cardiovascular system and air flow in the pulmonary system is vast, spanning simple one-dimensional models all the way to modern, patient-specific numerical simulations. Recent surveys~\cite{GrotbergJensen2004,Marsden2014} provide helpful entryways to this exciting field of study. 

Consistent with the guiding philosophy of this narrative, here we highlight a select few frontiers that are relatively unexplored and yet are potentially rich with new fluid mechanics. The first is in the context of the human brain. The role of fluid mechanics in transporting oxygen and nutrients to the brain has long been recognized. Indeed, strokes are fundamentally symptomatic of a failure of fluid transport in the blood vessels of the brain. However, it has recently become apparent that flow outside of the blood vessels, in the interstitial spaces in the brain, may be even more important for long-term brain function.

Bacyinski et al.~\cite{Bacyinski_etal_2017} lay a groundwork for studies of this problem, introducing the concept of a ``glymphatic'' system (a portmanteau that indicates a function like the lymphatic system but in the context of the brain's glial cells). An increasing body of empirical evidence suggests that flow through the brain is responsible for clearance of protein solutes and metabolic waste products that result from normal cell physiology. Dysfunction in that fluid transport has been at least indirectly associated with a wide range of disorders, including Alzheimer’s disease, strokes, migraine headaches, diabetes, and glaucoma~\cite{Bacyinski_etal_2017}. Factors that have been associated with dysfunction in the fluid transport include age, sleep deprivation, sleep apnea, and exercise (or lack thereof). 

The fluid dynamic mechanisms that could connect these various factors to brain function are unknown. In fact, there is not yet consensus on very basic aspects of the fluid mechanics. Where in the brain does the interstitial fluid flow, and how does it connect to cerebrospinal fluid? Does the fluid flow along the direction of blood flow and the corresponding arterial pressure pulse, or in the opposite direction? Is the fluid flow steady or unsteady? Answers to these questions and others will be essential to understand the role of fluid mechanics in brain function. For example, recall from the previous section that unsteady flow in the primitive heart tube amplified shear stresses by two orders of magnitude. 

Progress in this field will be challenged by obstacles similar to those we identified in the fluid mechanics of development. It is not yet possible to measure the glymphatic flows in real-time in humans. Animal models have enabled recent progress~\cite{Mestre_etal_2018}, and these combined with mechanical and numerical models are likely to be the most promising avenues in the near term. A key feature of the fluid dynamics to be resolved is the coupled-flow structure interactions that drive the glymphatic flow. Is this flow driven by pressure gradients in the interstitial region, or could deformation of blood vessels due to transient pressure inside of them be sufficient to create the interstitial flow? The dynamical coupling across the vessel walls is complicated by their nonlinear mechanical properties (which change with age) and the extremely complex geometry of the interstitial spaces. In this case, it is not obvious whether experiments or numerical simulations will be first to successfully capture the salient fluid mechanics.

The nascent stage of the aforementioned research stands in contrast to studies of airflow in the lungs, and yet it is here that we can find another intriguing frontier. The human pulmonary system is generally understood to comprise a stage of larger airways in which gases are transported advectively, followed by a bed of smaller structures in which gas exchange is diffusive. Inhalation and expiration change the direction of airflow through the system. 

Recent work suggests that this basic template is not followed by all vertebrates. Cieri and Farmer~\cite{CieriFarmer2016} show that in birds, crocodiles, and possibly also non-avian reptiles, the airflow is unidirectional. A series of aerodynamically actuated valves provide passive flow control to regulate gas exchange. In addition, even in the smallest airways, recent measurements suggest that advection may still remain important. These findings are intriguing for the potential consequences for gas exchange. If the residence time of a given parcel of air is limited by constant advection, how are the necessary fluxes of oxygen and carbon dioxide facilitated in these systems? Answers to questions like this can inspire new, high-throughput microfluidic technologies~\cite{StaplesMikel_2018}.

Our recurring challenge of limited observational powers appears in this context as well. Endoscopic cameras have been used to track aerosolized particles in the pulmonary system to achieve some estimates of these flows. However, we currently lack the spatiotemporal resolution to quantify the full air flow circuit in these animal systems. There is significant potential to learn new principles of flow control from these studies. It is also possible that more careful study in the context of simpler animal systems could lead to unexpected new insights into the human pulmonary system as well.

\subsection{\label{sec:level2f} Biological Communities}

\noindent\emph{\textbf{Landmarks:}  ``Marine microbes see a sea of gradients.'' Stocker. Science 338: 628--633 (2012)~\cite{Stocker2012}\\
\indent\indent\indent\indent\indent ``Flow and transport in regions with aquatic vegetation.'' Nepf. Annual Review of Fluid Mechanics 44: 123--142 (2012)~\cite{Nepf2012}
\\
\textbf{Frontier:} Resolving multi-scale fluid mechanics to understand emergent biological phenomena}
\vspace{5mm}

Biological systems rarely exist in isolation. From bacteria to animals to the plant life that constitutes the majority of biomass on earth, biological organisms function in communities. A consequence of this proximity between organisms is that the flow-structure interactions experienced by one of them can influence the fluid mechanics of its neighbors. Familiar manifestations of this phenomenon include the swimming of fish in schools and the formation flying of birds. But these represent a small subset of many ways in which communities of organisms are influenced by, and can themselves affect, fluid mechanics.

A common principle in analysis of community-scale fluid mechanics is that the dynamics at that scale can rarely be appreciated by a simple superposition of the fluid mechanics of each individual. Even at low Reynolds numbers, where the equations of motion become linear and can be analyzed using Green's functions, it is important to remember that the response of organisms to the surrounding flow can remain highly nonlinear. Organisms in swarms can take cues both from the surrounding fluid and from the behavior of adjacent neighbors to inform their own locomotion. Those decisions will in turn affect the surrounding fluid, creating a rich, coupled dynamical system~\cite{SinhuberOuellette2017}. The introduction of nonlinearity in the fluid dynamics, for example from inertial effects or nonlinear rheology, can add further complexity to the system.

This mantra of the whole as different from the sum of its parts can lead to surprising fluid mechanics, and this realization is just beginning to be appreciated as a frontier area of research in biological fluid dynamics. Stocker~\cite{Stocker2012} provides an excellent summary of how microbial communities in the ocean both exploit and create gradients that affect large-scale biogeochemistry. Rather than the microscale fluid mechanics simply averaging out, microscale gradients can accentuate differences in growth and behavior of individuals in the community, leading to even greater differences at macroscale. The formation of phytoplankton blooms, bacterial biofilms, and other aggregates can in many cases be traced to the amplification of a small initial fluid dynamic perturbation.

Interactions among organisms at the microscale is facilitated by viscous effects that dominate at low Reynolds numbers. In the limit of Stokes flow, the volume of surrounding fluid that is set into motion by a translating body becomes unbounded. Hence, the organisms in a low Reynolds number community are inextricably coupled by the fluid mechanics, even when their spacing becomes large.

One might expect that fluid dynamic interactions become less significant for communities at higher Reynolds numbers, wherein the region of the surrounding flow affected by each individual organism becomes compact. Somewhat surprisingly, significant community-scale effects can be still observed in this case as well, with a caveat. In order for the fluid dynamics to reflect length scales that significantly exceed the size of the individual organisms, a given parcel of fluid must experience interactions with multiple organisms in the community. A striking example of this was recently demonstrated~\cite{Houghton_etal_2018} in the context of the vertical migration of swimming plankton. An individual swimmer, with a characteristic Reynolds number of order 10 to 100 based on its swimming speed and body length, will propel a parcel of the surrounding fluid downward as it swims upward. That fluid parcel will continue moving downstream until its energy is dissipated by viscosity, or in the ocean, until buoyancy forces return the fluid to its equilibrium depth in the water.

However, if that parcel of fluid is subsequently forced downward by another swimmer in its path, the fluid can be propelled downstream further still. Ultimately, the distance over which the fluid is transported is then dictated not by the length scale of the individual swimmers, but rather by the vertical extent of the migrating community. An analogous effect is fundamental to ciliary transport: an individual cilium transports fluid over a distance comparable to its length, whereas the total distance of ciliary transport is limited only by the size of the entire community~\cite{Shapiro_etal_2014}.

These observations highlight the fact that an understanding of the fluid mechanics of an individual organism is typically insufficient to deduce the flow physics at community scale. At the same time, it is typically challenging to empirically model or to numerically simulate fluid dynamics across scales from individuals to communities~\cite{Katija2012}. The state of the art is often to coarse-grain the fluid mechanics at the scale of individuals, and to use the resulting parameterizations to study the dynamics at larger scales. While there have been some successes in this regard~\cite{Filella_etal_2018,Morrell_etal_2019}, it is often the case the local-scale fluid dynamics are too heterogeneous to be coarse-grained without losing essential physics. Models of collective vertical migration have been observed to underestimate fluid transport for this reason~\cite{Wilhelmus_etal_2019}. 

In other cases, there exist feedbacks between the community scale fluid mechanics and the local phenomena the lead to a closure problem akin to that in turbulence modeling. The context of plant canopies, as well summarized by Nepf~\cite{Nepf2012}, illustrates these dynamics clearly. Where plant canopies are sparse, the boundary layer of flow through the canopy may exhibit a classical logarithmic shape, with eddy scales dominated by the distance from the ground and the scale of the individual plants. In contrast, dense canopies can lead to an inflection point instability in the flow profile, with eddies forming at a length scale set by the momentum thickness of the mixing layer. In between these extremes, a spectrum of eddy sizes can coexist. The eddy field will influence the transport of oxygen, nutrients, and seeds, each of which can in turn affect the density of plants that can be supported. Hence, one cannot understand the dynamics of the individual plants without knowledge of fluid mechanics at canopy-scale; and concurrently, the structure of the canopy is determined by the fate of the individual plants.

Even putting the plant biology aside, inasmuch as the flow through the canopy depends on its structure, a complex fluid dynamic coupling arises when the plants can deform. The aerodynamic drag experienced by the plants located furthest upwind will be dictated by the incident wind conditions and any passive shape change that may reduce drag~\cite{Vogel1989,Alben_etal_2002}. A plant that is immediately downwind from the first will experience aerodynamic drag that depends on the incident wind conditions, the flow-induced shape of the plant located upwind, and any passive shape change that the second plant experiences. And so on. In sum, the fate of the downwind plants depends on all of the upwind flow-structure interactions. Given the aforementioned community-scale considerations for the transport of life-sustaining nutrients, the community-scale fluid mechanics may present an opportunity for optimization. Specifically, the flow-structure interactions of the plants located upwind can potentially be adapted to ensure the health of the community as whole, rather than the local plant alone.

It is interesting to note that this strategy of forsaking a ``greedy'' algorithm for each individual in a community has been adopted successfully in engineering applications such as wind farm design. In this case, it is known that if the performance of each wind turbine is maximized in isolation, the turbine located furthest upwind will perform well. However, the turbulent wake it creates will cause downwind turbine performance to suffer significantly. Efficiency reductions of up to 40 percent are observed in practice~\cite{Barthelmie_etal_2009}. By contrast, intentional reductions in the performance of the upwind turbine can be more than compensated by the improved energy generation of downwind turbines due to lower turbulence and wake losses~\cite{Howland_etal_2019}.

As we develop capabilities to resolve the multiscale fluid mechanics in communities, without losing essential physics in processes like coarse-graining, we will be able to more effectively include biological fluid dynamics in models of large-scale ecological processes. As noted above, the challenges in modeling these systems mirror those in turbulent flows. In the same way that large-eddy simulation leverages models of sub-grid-scale processes, it may be possible to pursue sub-grid-scale models of biological flow-structure interactions to enable simulation of community-scale dynamics.

The wind farm example is just one of many ways in which technology can be inspired by the fluid dynamics of communities. Robotic swarms have been proposed for search and rescue missions, drug delivery in the blood stream, and energy-efficient transportation fleets, among many other applications. Each of these contexts can potentially leverage solutions to analogous fluid mechanics challenges in nature, with benefits across health care, energy, and national security.

\section{\label{sec:level1c}Closing Remarks}

I would be remiss if I did not end this narrative with the same caveats offered at the beginning. The field of biological fluid dynamics is too rich, too diverse, and too dynamic to capture all of it in a brief discussion like this. In each of the preceding sections, large swaths of the field have been necessarily omitted. The analysis of locomotion neglected interesting fluid mechanics that arise at the interface between air and water, including the role of surface tension at small scales and of wave drag at larger scales~\cite{BushHu2006,Hu_etal_2010,MendelsonTechet2018}. As noted in the section on sensing, we have not discussed the role of fluid mechanics in hearing and phonation~\cite{KraneWei_2006,Mittal_etal_2013, Obrist2019}, functions that are essential in many predator-prey interactions and in mating. We described the role of fluid mechanics in conception during sexual reproduction, but other important fluid dynamic processes arise all the way through the culmination of pregnancy in the birthing process~\cite{Yaniv_etal_2003,Lehn_etal_2016}. Likewise, biological development occurs not only in the transition from embryonic to adult life forms, but also in renewal of mature biological machinery. Wound healing is one such example~\cite{Jiang_etal_2018}.

Research in biological fluid dynamics is arguably imbalanced toward animal life, despite the fact that the majority of life on earth is in the form of plants~\cite{BarOn_etal_2018}. This bias was reflected in the narrative on internal flows. Plants also rely on advective and diffusive transport of gases and dissolved nutrients, which highlights another important role for flow-structure interactions in deformable vessels. Finally, the concept of communities has many more fascinating manifestations, such as in bioconvection~\cite{Pedley2010} and bacterial symbioses~\cite{Nawroth_etal_2017}. We have only scratched the surface in our discussion of some of the commonly observed examples.

Notwithstanding these stipulations, the preceding narrative has hopefully given the reader a map of opportunities for progress in the field of biological fluid dynamics. From this analysis, we can also identify some recurring themes in terms of the need for new tools in experimental, analytical, and computational fluid dynamics.

The last significant advance in experimental fluid mechanics as applied to biological flows was arguably the introduction of laser-based flow velocimetry nearly 25 years ago~\cite{StamhuisVideler1995}. Recent efforts have aimed to expand this capability to three-component, three-dimensional measurements~\cite{Murphy_etal_2012,AdhikariLongmore2013,MendelsonTechet2018}; to derive pressure, force, and torque from biologically generated velocity fields~\cite{Dabiri_etal_2014,Lucas_etal_2017}; and to use ultrasound and magnetic resonance imaging to capture the salient flow kinematics using non-optical approaches~\cite{Kheradvar_etal_2010,Elkins_etal_2003}. A fundamental challenge in each of this efforts is that the enhanced measurement capabilities, especially three-dimensionality, often comes at the cost of a reduction in spatial or temporal resolution. In addition, the technological advances to date typically do not resolve the issue of line-of-sight access (optical or acoustic), which as we saw in the context of internal flows, is severely limiting progress. A notable exception for three-dimensional measurements is the advent of magnetic resonance velocimetry, which does not require optical or acoustic access~\cite{Elkins_etal_2003}. However, it does require instrumentation that is not commonly available to researchers in fluid mechanics, nor is it easily adaptable to many biological flows of interest outside of the context of human physiology. The temporal resolution of the resulting measurements is also still insufficient to measure flows that are aperiodic, as are most biological flows. 

If the next major advance in experimental biological fluid dynamics does not come from new measurement techniques, it could instead arise from new tools to affect animal behavior. A key limitation of basic science research in this area has been that we can currently only observe the fluid dynamics that organisms exhibit naturally. For example, because evolution by natural selection has not necessarily produced swimming and flying organisms that maximize swimming speed, efficiency, maneuverability, or stealth, we cannot rely on observations of voluntary locomotion to reveal the full envelope of optimal propulsion modes that is physically realizable.   Hypotheses regarding locomotion regimes that do not occur naturally have therefore only been accessible for testing via physical or computational models, with their attendant simplifications.

In a few notable cases, it has been possible to use MEMS technology to perturb animal behavior~\cite{SponbergDaniel2012,Sato_etal_2015}. If these capabilities can be expanded to enable robust control of locomotion, fluid transport, other biological functions, the range of new scientific questions that could be addressed in biological fluid mechanics would expand dramatically. These same tools that would benefit our understanding of natural biological function could also be leveraged for bio-inspired engineering technologies such as robotics.
	
Analytical and computational approaches have experienced dramatic advances due to more powerful hardware and new approaches to data analysis. Machine learning can potentially accelerate research tasks in biological fluid dynamics, such as object detection and tracking, and may also enable identification of correlations that are missed by manual inspection. An application with great potential in this regard is deducing the connection between fluid dynamic forces and gene expression during mechanotransduction. Current single-cell RNA sequencing studies can readily probe the expression of 10$^{4}$ genes in each of 10$^{4}$ disaggregated cells~\cite{Briggs_etal_2018}. Those orders of magnitude will continue to increase rapidly in the near future. It is likely that only through automated data analysis will it be possible to correlate gene expression to the complex spatiotemporal patterns of fluid dynamic forcing that could be effecting gene expression.

Numerical simulation is already enabling flows in complex geometries to be reproduced with high fidelity~\cite{SotiropoulosYang_2014,Marsden2014,vanRees_etal_2015}. A next frontier is coupling of flow simulation with the biology and chemistry of biological systems. Here, as in our discussion of community-scale fluid dynamics, there will be a need to develop clever methods to capture local, micro-scale physics concurrently with the fluid mechanics that rely on continuum approximations.  

In sum, the field of biological fluid dynamics continues to make great strides in advancing both biology and fluid mechanics. Yet, there remain many exciting opportunities for new ideas and new researchers to enter the field and to redefine its frontier. We will all benefit from their success, as new interventions are developed to sustain human health and the health of our planet. However, as we pursue these aims, let us not become overly utilitarian in our exploration of new physics. While I motivated this narrative on the basis of timely challenges facing society, I remain convinced that our innate curiosity about biological flows, often inspired simply by their unusual beauty, is likely to lead us toward the most interesting frontiers. 

\begin{acknowledgments}
The author thanks Margaret Byron and Lex Smits for organizing the minisymposium at the 2018 Annual Meeting of the American Physical Society Division of Fluid Dynamics, where many of these ideas were originally presented. Helpful feedback from Brooke Husic, Megan Leftwich, Clara O'Farrell, Robert Whittlesey, Don Webster, and the anonymous referees is also gratefully acknowledged.
\end{acknowledgments}

\bibliography{apssamp}

\providecommand{\noopsort}[1]{}\providecommand{\singleletter}[1]{#1}%
\begin{thebibliography}{116}%
\makeatletter
\providecommand \@ifxundefined [1]{%
 \@ifx{#1\undefined}
}%
\providecommand \@ifnum [1]{%
 \ifnum #1\expandafter \@firstoftwo
 \else \expandafter \@secondoftwo
 \fi
}%
\providecommand \@ifx [1]{%
 \ifx #1\expandafter \@firstoftwo
 \else \expandafter \@secondoftwo
 \fi
}%
\providecommand \natexlab [1]{#1}%
\providecommand \enquote  [1]{``#1''}%
\providecommand \bibnamefont  [1]{#1}%
\providecommand \bibfnamefont [1]{#1}%
\providecommand \citenamefont [1]{#1}%
\providecommand \href@noop [0]{\@secondoftwo}%
\providecommand \href [0]{\begingroup \@sanitize@url \@href}%
\providecommand \@href[1]{\@@startlink{#1}\@@href}%
\providecommand \@@href[1]{\endgroup#1\@@endlink}%
\providecommand \@sanitize@url [0]{\catcode `\\12\catcode `\$12\catcode
  `\&12\catcode `\#12\catcode `\^12\catcode `\_12\catcode `\%12\relax}%
\providecommand \@@startlink[1]{}%
\providecommand \@@endlink[0]{}%
\providecommand \url  [0]{\begingroup\@sanitize@url \@url }%
\providecommand \@url [1]{\endgroup\@href {#1}{\urlprefix }}%
\providecommand \urlprefix  [0]{URL }%
\providecommand \Eprint [0]{\href }%
\providecommand \doibase [0]{https://doi.org/}%
\providecommand \selectlanguage [0]{\@gobble}%
\providecommand \bibinfo  [0]{\@secondoftwo}%
\providecommand \bibfield  [0]{\@secondoftwo}%
\providecommand \translation [1]{[#1]}%
\providecommand \BibitemOpen [0]{}%
\providecommand \bibitemStop [0]{}%
\providecommand \bibitemNoStop [0]{.\EOS\space}%
\providecommand \EOS [0]{\spacefactor3000\relax}%
\providecommand \BibitemShut  [1]{\csname bibitem#1\endcsname}%
\let\auto@bib@innerbib\@empty
\bibitem [{WHO(2018)}]{WHO2018}%
  \BibitemOpen
  \href@noop {} {\emph {\bibinfo {title} {Global Health Estimates 2016: Deaths
  by Cause, Age, Sex, by Country and by Region, 2000-2016}}},\ \bibinfo {type}
  {Tech. Rep.}\ (\bibinfo  {institution} {World Health Organization, Geneva},\
  \bibinfo {year} {2018})\BibitemShut {NoStop}%
\bibitem [{Note1()}]{Note1}%
  \BibitemOpen
  \bibinfo {note} {Ischaemic heart disease; stroke; chronic obstructive
  pulmonary disease; lower respiratory infections; Alzheimer disease and other
  dementias; trachea, bronchus, and lung cancers; and diabetes
  mellitus}\BibitemShut {NoStop}%
\bibitem [{\citenamefont {Scharfman}\ \emph {et~al.}(2016)\citenamefont
  {Scharfman}, \citenamefont {Techet}, \citenamefont {Bush},\  and\
  \citenamefont {Bourouiba}}]{Scharfman_etal_2016}%
  \BibitemOpen
  \bibfield  {author} {\bibinfo {author} {\bibfnamefont {B.~E.}\ \bibnamefont
  {Scharfman}}, \bibinfo {author} {\bibfnamefont {A.~H.}\ \bibnamefont
  {Techet}}, \bibinfo {author} {\bibfnamefont {J.~W.~M.}\ \bibnamefont {Bush}},
   and\ \bibinfo {author} {\bibfnamefont {L.}~\bibnamefont {Bourouiba}},\
  }\bibfield  {title} {\bibinfo {title} {Visualization of sneeze ejecta: steps
  of fluid fragmentation leading to respiratory droplets},\ }\href@noop {}
  {\bibfield  {journal} {\bibinfo  {journal} {Experiments in Fluids}\ }\textbf
  {\bibinfo {volume} {57}},\ \bibinfo {pages} {24} (\bibinfo {year}
  {2016})}\BibitemShut {NoStop}%
\bibitem [{\citenamefont {Kim}\ \emph {et~al.}(2019)\citenamefont {Kim},
  \citenamefont {Park}, \citenamefont {Gruszewski}, \citenamefont {III},\  and\
  \citenamefont {Jung}}]{Kim_etal_2019}%
  \BibitemOpen
  \bibfield  {author} {\bibinfo {author} {\bibfnamefont {S.}~\bibnamefont
  {Kim}}, \bibinfo {author} {\bibfnamefont {H.}~\bibnamefont {Park}}, \bibinfo
  {author} {\bibfnamefont {H.}~\bibnamefont {Gruszewski}}, \bibinfo {author}
  {\bibfnamefont {D.~G.~S.}\ \bibnamefont {III}},  and\ \bibinfo {author}
  {\bibfnamefont {S.}~\bibnamefont {Jung}},\ }\bibfield  {title} {\bibinfo
  {title} {Vortex-induced dispersal of a plant pathogen by raindrop impact},\
  }\href@noop {} {\bibfield  {journal} {\bibinfo  {journal} {Proceedings of the
  National Academy of Sciences}\ }\textbf {\bibinfo {volume} {116}},\ \bibinfo
  {pages} {4917} (\bibinfo {year} {2019})}\BibitemShut {NoStop}%
\bibitem [{\citenamefont {Koumoutsakos}\ \emph {et~al.}(2013)\citenamefont
  {Koumoutsakos}, \citenamefont {Pivkin},\  and\ \citenamefont
  {Milde}}]{Koumoutsakos_etal_2013}%
  \BibitemOpen
  \bibfield  {author} {\bibinfo {author} {\bibfnamefont {P.}~\bibnamefont
  {Koumoutsakos}}, \bibinfo {author} {\bibfnamefont {I.}~\bibnamefont
  {Pivkin}},  and\ \bibinfo {author} {\bibfnamefont {F.}~\bibnamefont
  {Milde}},\ }\bibfield  {title} {\bibinfo {title} {The fluid mechanics of
  cancer and its therapy},\ }\href@noop {} {\bibfield  {journal} {\bibinfo
  {journal} {Annual Review of Fluid Mechanics}\ }\textbf {\bibinfo {volume}
  {45}},\ \bibinfo {pages} {325} (\bibinfo {year} {2013})}\BibitemShut
  {NoStop}%
\bibitem [{\citenamefont {Bacyinski}\ \emph {et~al.}(2017)\citenamefont
  {Bacyinski}, \citenamefont {Xu}, \citenamefont {Wang},\  and\ \citenamefont
  {Hu}}]{Bacyinski_etal_2017}%
  \BibitemOpen
  \bibfield  {author} {\bibinfo {author} {\bibfnamefont {A.}~\bibnamefont
  {Bacyinski}}, \bibinfo {author} {\bibfnamefont {M.}~\bibnamefont {Xu}},
  \bibinfo {author} {\bibfnamefont {W.}~\bibnamefont {Wang}},  and\ \bibinfo
  {author} {\bibfnamefont {J.}~\bibnamefont {Hu}},\ }\bibfield  {title}
  {\bibinfo {title} {The paravascular pathway for brain waste clearance:
  current understanding, significance and controversy},\ }\href@noop {}
  {\bibfield  {journal} {\bibinfo  {journal} {Frontiers in Neuroanatomy}\
  }\textbf {\bibinfo {volume} {11}},\ \bibinfo {pages} {101} (\bibinfo {year}
  {2017})}\BibitemShut {NoStop}%
\bibitem [{\citenamefont {Mayer}\ \emph {et~al.}(2014)\citenamefont {Mayer},
  \citenamefont {Knight}, \citenamefont {Mazmanian}, \citenamefont {Cryan},\
  and\ \citenamefont {Tillisch}}]{Mayer_etal_2014}%
  \BibitemOpen
  \bibfield  {author} {\bibinfo {author} {\bibfnamefont {E.~A.}\ \bibnamefont
  {Mayer}}, \bibinfo {author} {\bibfnamefont {R.}~\bibnamefont {Knight}},
  \bibinfo {author} {\bibfnamefont {S.~K.}\ \bibnamefont {Mazmanian}}, \bibinfo
  {author} {\bibfnamefont {J.~F.}\ \bibnamefont {Cryan}},  and\ \bibinfo
  {author} {\bibfnamefont {K.}~\bibnamefont {Tillisch}},\ }\bibfield  {title}
  {\bibinfo {title} {Gut microbes and the brain: paradigm shift in
  neuroscience},\ }\href@noop {} {\bibfield  {journal} {\bibinfo  {journal}
  {Journal of Neuroscience}\ }\textbf {\bibinfo {volume} {34}},\ \bibinfo
  {pages} {15490} (\bibinfo {year} {2014})}\BibitemShut {NoStop}%
\bibitem [{\citenamefont {Bloom}\ \emph {et~al.}(2012)\citenamefont {Bloom},
  \citenamefont {Cafiero}, \citenamefont {Jan-Llopis}, \citenamefont
  {Abrahams-Gessel}, \citenamefont {Bloom}, \citenamefont {Fathima},
  \citenamefont {Feigl}, \citenamefont {Gaziano}, \citenamefont {Hamandi},\
  and\ \citenamefont {Mowafi}}]{Bloom_etal_2012}%
  \BibitemOpen
  \bibfield  {author} {\bibinfo {author} {\bibfnamefont {D.~E.}\ \bibnamefont
  {Bloom}}, \bibinfo {author} {\bibfnamefont {E.}~\bibnamefont {Cafiero}},
  \bibinfo {author} {\bibfnamefont {E.}~\bibnamefont {Jan-Llopis}}, \bibinfo
  {author} {\bibfnamefont {S.}~\bibnamefont {Abrahams-Gessel}}, \bibinfo
  {author} {\bibfnamefont {L.~R.}\ \bibnamefont {Bloom}}, \bibinfo {author}
  {\bibfnamefont {S.}~\bibnamefont {Fathima}}, \bibinfo {author} {\bibfnamefont
  {A.~B.}\ \bibnamefont {Feigl}}, \bibinfo {author} {\bibfnamefont
  {T.}~\bibnamefont {Gaziano}}, \bibinfo {author} {\bibfnamefont
  {A.}~\bibnamefont {Hamandi}},  and\ \bibinfo {author} {\bibfnamefont
  {M.}~\bibnamefont {Mowafi}},\ }\href@noop {} {\emph {\bibinfo {title} {The
  global economic burden ofnoncommunicable diseases}}},\ \bibinfo {type} {Tech.
  Rep.}\ (\bibinfo  {institution} {Program on the Global Demography of Aging,
  Geneva, Switzerland},\ \bibinfo {year} {2012})\BibitemShut {NoStop}%
\bibitem [{EIA(2017)}]{EIA2017}%
  \BibitemOpen
  \href@noop {} {\emph {\bibinfo {title} {Global Transportation Energy
  Consumption: Examination of Scenarios to 2040 using ITEDD}}},\ \bibinfo
  {type} {Tech. Rep.}\ (\bibinfo  {institution} {U. S. Energy Information
  Administration},\ \bibinfo {year} {2017})\BibitemShut {NoStop}%
\bibitem [{\citenamefont {Gemmell}\ \emph {et~al.}(2013)\citenamefont
  {Gemmell}, \citenamefont {Costello}, \citenamefont {Colin}, \citenamefont
  {Stewart}, \citenamefont {Dabiri}, \citenamefont {Tafti},\  and\
  \citenamefont {Priya}}]{Gemmell_etal_2013}%
  \BibitemOpen
  \bibfield  {author} {\bibinfo {author} {\bibfnamefont {B.~J.}\ \bibnamefont
  {Gemmell}}, \bibinfo {author} {\bibfnamefont {J.~H.}\ \bibnamefont
  {Costello}}, \bibinfo {author} {\bibfnamefont {S.~P.}\ \bibnamefont {Colin}},
  \bibinfo {author} {\bibfnamefont {C.}~\bibnamefont {Stewart}}, \bibinfo
  {author} {\bibfnamefont {J.~O.}\ \bibnamefont {Dabiri}}, \bibinfo {author}
  {\bibfnamefont {D.}~\bibnamefont {Tafti}},  and\ \bibinfo {author}
  {\bibfnamefont {S.}~\bibnamefont {Priya}},\ }\bibfield  {title} {\bibinfo
  {title} {Passive energy recapture in jellyfish contributes to propulsive
  advantage over other metazoans},\ }\href@noop {} {\bibfield  {journal}
  {\bibinfo  {journal} {Proceedings of the National Academy of Sciences}\
  }\textbf {\bibinfo {volume} {110}},\ \bibinfo {pages} {17904} (\bibinfo
  {year} {2013})}\BibitemShut {NoStop}%
\bibitem [{Pri(2017)}]{Prius2017}%
  \BibitemOpen
  \href@noop {} {\emph {\bibinfo {title} {Fuel Economy of the 2017 Toyota Prius
  Eco}}},\ \bibinfo {type} {www.fueleconomy.gov}\ (\bibinfo  {institution} {U.
  S. Department of Energy},\ \bibinfo {year} {2017})\BibitemShut {NoStop}%
\bibitem [{\citenamefont {Peeters}\ \emph {et~al.}(2005)\citenamefont
  {Peeters}, \citenamefont {Middel},\  and\ \citenamefont
  {Hoolhorst}}]{Peeters_etal_2005}%
  \BibitemOpen
  \bibfield  {author} {\bibinfo {author} {\bibfnamefont {P.~M.}\ \bibnamefont
  {Peeters}}, \bibinfo {author} {\bibfnamefont {J.}~\bibnamefont {Middel}},
  and\ \bibinfo {author} {\bibfnamefont {A.}~\bibnamefont {Hoolhorst}},\
  }\href@noop {} {\emph {\bibinfo {title} {Fuel efficiency of commercial
  aircraft: an overview of historical and future trends}}},\ \bibinfo {type}
  {{Technical Report NLR-CR-2005-669}}\ (\bibinfo  {institution} {National
  Aerospace Laboratory, The Netherlands},\ \bibinfo {year} {2005})\BibitemShut
  {NoStop}%
\bibitem [{\citenamefont {Hornung}(2006)}]{Hornung2006}%
  \BibitemOpen
  \bibfield  {author} {\bibinfo {author} {\bibfnamefont {H.~G.}\ \bibnamefont
  {Hornung}},\ }\href@noop {} {\emph {\bibinfo {title} {Dimensional Analysis:
  Examples of the Use of Symmetry}}}\ (\bibinfo  {publisher} {Dover
  Publications},\ \bibinfo {year} {2006})\BibitemShut {NoStop}%
\bibitem [{\citenamefont {Olivier}\ \emph {et~al.}(2016)\citenamefont
  {Olivier}, \citenamefont {Janssens-Maenhout}, \citenamefont {Muntean},\  and\
  \citenamefont {Peters}}]{Olivier_etal_2016}%
  \BibitemOpen
  \bibfield  {author} {\bibinfo {author} {\bibfnamefont {J.~G.~J.}\
  \bibnamefont {Olivier}}, \bibinfo {author} {\bibfnamefont {G.}~\bibnamefont
  {Janssens-Maenhout}}, \bibinfo {author} {\bibfnamefont {M.}~\bibnamefont
  {Muntean}},  and\ \bibinfo {author} {\bibfnamefont {J.~A. H.~W.}\
  \bibnamefont {Peters}},\ }\href@noop {} {\emph {\bibinfo {title} {Trends in
  global CO2 emissions: 2016 Report}}},\ \bibinfo {type} {{Technical Report
  JRC103425}}\ (\bibinfo  {institution} {European Commission, Joint Research
  Centre (JRC), Directorate C - Energy, Transport and Climate},\ \bibinfo
  {year} {2016})\BibitemShut {NoStop}%
\bibitem [{IPC(2018)}]{IPCC2018}%
  \BibitemOpen
  \href@noop {} {\emph {\bibinfo {title} {Global warming of 1.5 degrees C: An
  IPCC special report on the impacts of global warming of 1.5 degrees C above
  pre-industrial levels and related global greenhouse gas emission pathways, in
  the context of strengthening the global response to the threat of climate
  change, sustainable development, and efforts to eradicate poverty}}},\
  \bibinfo {type} {Tech. Rep.}\ (\bibinfo  {institution} {Intergovernmental
  Panel on Climate Change},\ \bibinfo {year} {2018})\BibitemShut {NoStop}%
\bibitem [{\citenamefont {Fish}\ \emph {et~al.}(2011)\citenamefont {Fish},
  \citenamefont {Weber}, \citenamefont {Murray},\  and\ \citenamefont
  {Howle}}]{Fish_etal_2011}%
  \BibitemOpen
  \bibfield  {author} {\bibinfo {author} {\bibfnamefont {F.~E.}\ \bibnamefont
  {Fish}}, \bibinfo {author} {\bibfnamefont {P.~W.}\ \bibnamefont {Weber}},
  \bibinfo {author} {\bibfnamefont {M.~M.}\ \bibnamefont {Murray}},  and\
  \bibinfo {author} {\bibfnamefont {L.~E.}\ \bibnamefont {Howle}},\ }\bibfield
  {title} {\bibinfo {title} {Marine applications of the biomimetic humpback
  whale flipper},\ }\href@noop {} {\bibfield  {journal} {\bibinfo  {journal}
  {Marine Technology Society Journal}\ }\textbf {\bibinfo {volume} {45}},\
  \bibinfo {pages} {198} (\bibinfo {year} {2011})}\BibitemShut {NoStop}%
\bibitem [{\citenamefont {Whittlesey}\ \emph {et~al.}(2010)\citenamefont
  {Whittlesey}, \citenamefont {Liska},\  and\ \citenamefont
  {Dabiri}}]{Whittlesey_etal_2010}%
  \BibitemOpen
  \bibfield  {author} {\bibinfo {author} {\bibfnamefont {R.~W.}\ \bibnamefont
  {Whittlesey}}, \bibinfo {author} {\bibfnamefont {S.~C.}\ \bibnamefont
  {Liska}},  and\ \bibinfo {author} {\bibfnamefont {J.~O.}\ \bibnamefont
  {Dabiri}},\ }\bibfield  {title} {\bibinfo {title} {Fish schooling as a basis
  for vertical-axis wind turbine farm design},\ }\href@noop {} {\bibfield
  {journal} {\bibinfo  {journal} {Bioinspiration and Biomimetics}\ }\textbf
  {\bibinfo {volume} {5}},\ \bibinfo {pages} {035005} (\bibinfo {year}
  {2010})}\BibitemShut {NoStop}%
\bibitem [{\citenamefont {Vogel}(1989)}]{Vogel1989}%
  \BibitemOpen
  \bibfield  {author} {\bibinfo {author} {\bibfnamefont {S.}~\bibnamefont
  {Vogel}},\ }\bibfield  {title} {\bibinfo {title} {Drag and reconfiguration of
  broad leaves in high winds},\ }\href@noop {} {\bibfield  {journal} {\bibinfo
  {journal} {Journal of Experimental Botany}\ }\textbf {\bibinfo {volume}
  {40}},\ \bibinfo {pages} {941} (\bibinfo {year} {1989})}\BibitemShut
  {NoStop}%
\bibitem [{\citenamefont {Taylor}\ \emph {et~al.}(2003)\citenamefont {Taylor},
  \citenamefont {Nudds},\  and\ \citenamefont {Thomas}}]{Taylor_etal_2003}%
  \BibitemOpen
  \bibfield  {author} {\bibinfo {author} {\bibfnamefont {G.~K.}\ \bibnamefont
  {Taylor}}, \bibinfo {author} {\bibfnamefont {R.~L.}\ \bibnamefont {Nudds}},
  and\ \bibinfo {author} {\bibfnamefont {A.~L.~R.}\ \bibnamefont {Thomas}},\
  }\bibfield  {title} {\bibinfo {title} {Flying and swimming animals cruise at
  a strouhal number tuned for high power efficiency},\ }\href@noop {}
  {\bibfield  {journal} {\bibinfo  {journal} {Nature}\ }\textbf {\bibinfo
  {volume} {425}},\ \bibinfo {pages} {707} (\bibinfo {year}
  {2003})}\BibitemShut {NoStop}%
\bibitem [{\citenamefont {Triantafyllou}\ \emph {et~al.}(1991)\citenamefont
  {Triantafyllou}, \citenamefont {Triantafyllou},\  and\ \citenamefont
  {Gopalkrishnan}}]{Triantafyllou_etal_1991}%
  \BibitemOpen
  \bibfield  {author} {\bibinfo {author} {\bibfnamefont {M.~S.}\ \bibnamefont
  {Triantafyllou}}, \bibinfo {author} {\bibfnamefont {G.~S.}\ \bibnamefont
  {Triantafyllou}},  and\ \bibinfo {author} {\bibfnamefont {R.}~\bibnamefont
  {Gopalkrishnan}},\ }\bibfield  {title} {\bibinfo {title} {Wake mechanics for
  thrust generation in oscillating foils},\ }\href@noop {} {\bibfield
  {journal} {\bibinfo  {journal} {Physics of Fluids A: Fluid Dynamics}\
  }\textbf {\bibinfo {volume} {3}},\ \bibinfo {pages} {2835} (\bibinfo {year}
  {1991})}\BibitemShut {NoStop}%
\bibitem [{\citenamefont {Floryan}\ \emph {et~al.}(2018)\citenamefont
  {Floryan}, \citenamefont {Buren},\  and\ \citenamefont
  {Smits}}]{Floryan_etal_2018}%
  \BibitemOpen
  \bibfield  {author} {\bibinfo {author} {\bibfnamefont {D.}~\bibnamefont
  {Floryan}}, \bibinfo {author} {\bibfnamefont {T.~V.}\ \bibnamefont {Buren}},
  and\ \bibinfo {author} {\bibfnamefont {A.~J.}\ \bibnamefont {Smits}},\
  }\bibfield  {title} {\bibinfo {title} {Efficient cruising for swimming and
  flying animals is dictated by fluid drag},\ }\href@noop {} {\bibfield
  {journal} {\bibinfo  {journal} {Proceedings of the National Academy of
  Sciences}\ }\textbf {\bibinfo {volume} {115}},\ \bibinfo {pages} {8116}
  (\bibinfo {year} {2018})}\BibitemShut {NoStop}%
\bibitem [{\citenamefont {Lentink}\ \emph {et~al.}(2009)\citenamefont
  {Lentink}, \citenamefont {Dickson}, \citenamefont {van Leeuwen},\  and\
  \citenamefont {Dickinson}}]{Lentink_etal_2009}%
  \BibitemOpen
  \bibfield  {author} {\bibinfo {author} {\bibfnamefont {D.}~\bibnamefont
  {Lentink}}, \bibinfo {author} {\bibfnamefont {W.~B.}\ \bibnamefont
  {Dickson}}, \bibinfo {author} {\bibfnamefont {J.~L.}\ \bibnamefont {van
  Leeuwen}},  and\ \bibinfo {author} {\bibfnamefont {M.~H.}\ \bibnamefont
  {Dickinson}},\ }\bibfield  {title} {\bibinfo {title} {Leading-edge vortices
  elevate lift of autorotating plant seeds},\ }\href@noop {} {\bibfield
  {journal} {\bibinfo  {journal} {Science}\ }\textbf {\bibinfo {volume}
  {324}},\ \bibinfo {pages} {1438} (\bibinfo {year} {2009})}\BibitemShut
  {NoStop}%
\bibitem [{\citenamefont {Muijres}\ \emph {et~al.}(2008)\citenamefont
  {Muijres}, \citenamefont {Johansson}, \citenamefont {Barfield}, \citenamefont
  {Wolf}, \citenamefont {Spedding},\  and\ \citenamefont
  {Hedenstrom}}]{Muijres_etal_2008}%
  \BibitemOpen
  \bibfield  {author} {\bibinfo {author} {\bibfnamefont {F.~T.}\ \bibnamefont
  {Muijres}}, \bibinfo {author} {\bibfnamefont {L.~C.}\ \bibnamefont
  {Johansson}}, \bibinfo {author} {\bibfnamefont {R.}~\bibnamefont {Barfield}},
  \bibinfo {author} {\bibfnamefont {M.}~\bibnamefont {Wolf}}, \bibinfo {author}
  {\bibfnamefont {G.~R.}\ \bibnamefont {Spedding}},  and\ \bibinfo {author}
  {\bibfnamefont {A.}~\bibnamefont {Hedenstrom}},\ }\bibfield  {title}
  {\bibinfo {title} {Leading-edge vortex improves lift in slow-flying bats},\
  }\href@noop {} {\bibfield  {journal} {\bibinfo  {journal} {Science}\ }\textbf
  {\bibinfo {volume} {319}},\ \bibinfo {pages} {1250} (\bibinfo {year}
  {2008})}\BibitemShut {NoStop}%
\bibitem [{\citenamefont {Hubel}\ \emph {et~al.}(2016)\citenamefont {Hubel},
  \citenamefont {Hristov}, \citenamefont {Swartz},\  and\ \citenamefont
  {Breuer}}]{Hubel_etal_2016}%
  \BibitemOpen
  \bibfield  {author} {\bibinfo {author} {\bibfnamefont {T.~Y.}\ \bibnamefont
  {Hubel}}, \bibinfo {author} {\bibfnamefont {N.~I.}\ \bibnamefont {Hristov}},
  \bibinfo {author} {\bibfnamefont {S.~M.}\ \bibnamefont {Swartz}},  and\
  \bibinfo {author} {\bibfnamefont {K.~S.}\ \bibnamefont {Breuer}},\ }\bibfield
   {title} {\bibinfo {title} {Wake structure and kinematics in two
  insectivorous bats},\ }\href@noop {} {\bibfield  {journal} {\bibinfo
  {journal} {Philosophical Transactions of the Royal Society B}\ }\textbf
  {\bibinfo {volume} {371}},\ \bibinfo {pages} {20150385} (\bibinfo {year}
  {2016})}\BibitemShut {NoStop}%
\bibitem [{\citenamefont {Gharib}\ \emph {et~al.}(1998)\citenamefont {Gharib},
  \citenamefont {Rambod},\  and\ \citenamefont {Shariff}}]{Gharib_etal_1998}%
  \BibitemOpen
  \bibfield  {author} {\bibinfo {author} {\bibfnamefont {M.}~\bibnamefont
  {Gharib}}, \bibinfo {author} {\bibfnamefont {E.}~\bibnamefont {Rambod}},
  and\ \bibinfo {author} {\bibfnamefont {K.}~\bibnamefont {Shariff}},\
  }\bibfield  {title} {\bibinfo {title} {A universal time scale for vortex ring
  formation},\ }\href@noop {} {\bibfield  {journal} {\bibinfo  {journal}
  {Journal of Fluid Mechanics}\ }\textbf {\bibinfo {volume} {360}},\ \bibinfo
  {pages} {121} (\bibinfo {year} {1998})}\BibitemShut {NoStop}%
\bibitem [{\citenamefont {Krueger}\  and\ \citenamefont
  {Gharib}(2003)}]{Krueger_Gharib_2003}%
  \BibitemOpen
  \bibfield  {author} {\bibinfo {author} {\bibfnamefont {P.~S.}\ \bibnamefont
  {Krueger}} and\ \bibinfo {author} {\bibfnamefont {M.}~\bibnamefont
  {Gharib}},\ }\bibfield  {title} {\bibinfo {title} {The significance of vortex
  ring formation to the impulse and thrust of a starting jet},\ }\href@noop {}
  {\bibfield  {journal} {\bibinfo  {journal} {Physics of Fluids}\ }\textbf
  {\bibinfo {volume} {15}},\ \bibinfo {pages} {1271} (\bibinfo {year}
  {2003})}\BibitemShut {NoStop}%
\bibitem [{\citenamefont {Saffman}(1992)}]{Saffman1992}%
  \BibitemOpen
  \bibfield  {author} {\bibinfo {author} {\bibfnamefont {P.~G.}\ \bibnamefont
  {Saffman}},\ }\href@noop {} {\emph {\bibinfo {title} {Vortex Dynamics}}}\
  (\bibinfo  {publisher} {Cambridge Monographs on Mechanics},\ \bibinfo {year}
  {1992})\BibitemShut {NoStop}%
\bibitem [{\citenamefont {Rosenfeld}\ \emph {et~al.}(1998)\citenamefont
  {Rosenfeld}, \citenamefont {Rambod},\  and\ \citenamefont
  {Gharib}}]{Rosenfeld_etal_1998}%
  \BibitemOpen
  \bibfield  {author} {\bibinfo {author} {\bibfnamefont {M.}~\bibnamefont
  {Rosenfeld}}, \bibinfo {author} {\bibfnamefont {E.}~\bibnamefont {Rambod}},
  and\ \bibinfo {author} {\bibfnamefont {M.}~\bibnamefont {Gharib}},\
  }\bibfield  {title} {\bibinfo {title} {Circulation and formation number of
  laminar vortex rings},\ }\href@noop {} {\bibfield  {journal} {\bibinfo
  {journal} {Journal of Fluid Mechanics}\ }\textbf {\bibinfo {volume} {376}},\
  \bibinfo {pages} {297} (\bibinfo {year} {1998})}\BibitemShut {NoStop}%
\bibitem [{\citenamefont {O'Farrell}\  and\ \citenamefont
  {Dabiri}(2014)}]{OFarrellDabiri2014}%
  \BibitemOpen
  \bibfield  {author} {\bibinfo {author} {\bibfnamefont {C.}~\bibnamefont
  {O'Farrell}} and\ \bibinfo {author} {\bibfnamefont {J.~O.}\ \bibnamefont
  {Dabiri}},\ }\bibfield  {title} {\bibinfo {title} {Nested contour dynamics
  models for axisymmetric vortex rings and vortex wakes},\ }\href@noop {}
  {\bibfield  {journal} {\bibinfo  {journal} {Journal of Fluid Mechanics}\
  }\textbf {\bibinfo {volume} {748}},\ \bibinfo {pages} {521} (\bibinfo {year}
  {2014})}\BibitemShut {NoStop}%
\bibitem [{\citenamefont {Kelvin}(1880)}]{Kelvin1880}%
  \BibitemOpen
  \bibfield  {author} {\bibinfo {author} {\bibfnamefont {L.}~\bibnamefont
  {Kelvin}},\ }\bibfield  {title} {\bibinfo {title} {Vortex statics},\
  }\href@noop {} {\bibfield  {journal} {\bibinfo  {journal} {Philosophical
  Magazine}\ }\textbf {\bibinfo {volume} {10}},\ \bibinfo {pages} {97}
  (\bibinfo {year} {1880})}\BibitemShut {NoStop}%
\bibitem [{\citenamefont {Benjamin}(1976)}]{Benjamin1976}%
  \BibitemOpen
  \bibfield  {author} {\bibinfo {author} {\bibfnamefont {T.~B.}\ \bibnamefont
  {Benjamin}},\ }\bibinfo {title} {The alliance of practical and analytical
  insights into the nonlinear problems of fluid mechanics},\ in\ \href@noop {}
  {\emph {\bibinfo {booktitle} {Applications of Methods of Functional Analysis
  to Problems in Mechanics}}},\ \bibinfo {editor} {edited by\ \bibinfo {editor}
  {\bibfnamefont {P.}~\bibnamefont {Germain}} and\ \bibinfo {editor}
  {\bibfnamefont {B.}~\bibnamefont {Nayroles}}}\ (\bibinfo  {publisher}
  {Springer},\ \bibinfo {address} {New York},\ \bibinfo {year} {1976})\ pp.\
  \bibinfo {pages} {8--28}\BibitemShut {NoStop}%
\bibitem [{\citenamefont {Dabiri}(2009)}]{Dabiri2009}%
  \BibitemOpen
  \bibfield  {author} {\bibinfo {author} {\bibfnamefont {J.~O.}\ \bibnamefont
  {Dabiri}},\ }\bibfield  {title} {\bibinfo {title} {Optimal vortex formation
  as a unifying principle in biological propulsion},\ }\href@noop {} {\bibfield
   {journal} {\bibinfo  {journal} {Annual Review of Fluid Mechanics}\ }\textbf
  {\bibinfo {volume} {41}},\ \bibinfo {pages} {17} (\bibinfo {year}
  {2009})}\BibitemShut {NoStop}%
\bibitem [{\citenamefont {Tytell}\ \emph {et~al.}(2010)\citenamefont {Tytell},
  \citenamefont {Hsu}, \citenamefont {Williams}, \citenamefont {Cohen},\  and\
  \citenamefont {Fauci}}]{Tytell_etal_2010}%
  \BibitemOpen
  \bibfield  {author} {\bibinfo {author} {\bibfnamefont {E.~D.}\ \bibnamefont
  {Tytell}}, \bibinfo {author} {\bibfnamefont {C.~Y.}\ \bibnamefont {Hsu}},
  \bibinfo {author} {\bibfnamefont {T.~L.}\ \bibnamefont {Williams}}, \bibinfo
  {author} {\bibfnamefont {A.~H.}\ \bibnamefont {Cohen}},  and\ \bibinfo
  {author} {\bibfnamefont {L.~J.}\ \bibnamefont {Fauci}},\ }\bibfield  {title}
  {\bibinfo {title} {Interactions between internal forces, body stiffness, and
  fluid environment in a neuromechanical model of lamprey swimming},\
  }\href@noop {} {\bibfield  {journal} {\bibinfo  {journal} {Proceedings of the
  National Academy of Sciences}\ }\textbf {\bibinfo {volume} {107}},\ \bibinfo
  {pages} {19832} (\bibinfo {year} {2010})}\BibitemShut {NoStop}%
\bibitem [{\citenamefont {Eldredge}\ \emph {et~al.}(2010)\citenamefont
  {Eldredge}, \citenamefont {Toomey},\  and\ \citenamefont
  {Medina}}]{Eldredge_etal_2010}%
  \BibitemOpen
  \bibfield  {author} {\bibinfo {author} {\bibfnamefont {J.~D.}\ \bibnamefont
  {Eldredge}}, \bibinfo {author} {\bibfnamefont {J.}~\bibnamefont {Toomey}},
  and\ \bibinfo {author} {\bibfnamefont {A.}~\bibnamefont {Medina}},\
  }\bibfield  {title} {\bibinfo {title} {On the roles of chord-wise flexibility
  in a flapping wing with hovering kinematics},\ }\href@noop {} {\bibfield
  {journal} {\bibinfo  {journal} {Journal of Fluid Mechanics}\ }\textbf
  {\bibinfo {volume} {659}},\ \bibinfo {pages} {94} (\bibinfo {year}
  {2010})}\BibitemShut {NoStop}%
\bibitem [{\citenamefont {Biewener}(2003)}]{Biewener2003}%
  \BibitemOpen
  \bibfield  {author} {\bibinfo {author} {\bibfnamefont {A.~A.}\ \bibnamefont
  {Biewener}},\ }\href@noop {} {\emph {\bibinfo {title} {Animal Locomotion}}}\
  (\bibinfo  {publisher} {Oxford Animal Biology Series},\ \bibinfo {year}
  {2003})\BibitemShut {NoStop}%
\bibitem [{\citenamefont {Leftwich}\ \emph {et~al.}(2012)\citenamefont
  {Leftwich}, \citenamefont {Tytell}, \citenamefont {Cohen},\  and\
  \citenamefont {Smits}}]{Leftwich_etal_2012}%
  \BibitemOpen
  \bibfield  {author} {\bibinfo {author} {\bibfnamefont {M.~C.}\ \bibnamefont
  {Leftwich}}, \bibinfo {author} {\bibfnamefont {E.~D.}\ \bibnamefont
  {Tytell}}, \bibinfo {author} {\bibfnamefont {A.~H.}\ \bibnamefont {Cohen}},
  and\ \bibinfo {author} {\bibfnamefont {A.~J.}\ \bibnamefont {Smits}},\
  }\bibfield  {title} {\bibinfo {title} {Wake structures behind a swimming
  robotic lamprey with a passively flexible tail},\ }\href@noop {} {\bibfield
  {journal} {\bibinfo  {journal} {Journal of Experimental Biology}\ }\textbf
  {\bibinfo {volume} {215}},\ \bibinfo {pages} {416} (\bibinfo {year}
  {2012})}\BibitemShut {NoStop}%
\bibitem [{\citenamefont {Seuront}\ \emph {et~al.}(2004)\citenamefont
  {Seuront}, \citenamefont {Schmitt}, \citenamefont {Souissi}, \citenamefont
  {Brewer},\  and\ \citenamefont {Strickler}}]{Seuront_etal_2004}%
  \BibitemOpen
  \bibfield  {author} {\bibinfo {author} {\bibfnamefont {L.}~\bibnamefont
  {Seuront}}, \bibinfo {author} {\bibfnamefont {F.~G.}\ \bibnamefont
  {Schmitt}}, \bibinfo {author} {\bibfnamefont {S.}~\bibnamefont {Souissi}},
  \bibinfo {author} {\bibfnamefont {M.}~\bibnamefont {Brewer}},  and\ \bibinfo
  {author} {\bibfnamefont {J.}~\bibnamefont {Strickler}},\ }\bibfield  {title}
  {\bibinfo {title} {From random walk to multifractal random walk in
  zooplankton swimming behaviour},\ }\href@noop {} {\bibfield  {journal}
  {\bibinfo  {journal} {Zoological Studies}\ }\textbf {\bibinfo {volume}
  {43}},\ \bibinfo {pages} {498} (\bibinfo {year} {2004})}\BibitemShut
  {NoStop}%
\bibitem [{\citenamefont {Dabiri}\ \emph {et~al.}(2019)\citenamefont {Dabiri},
  \citenamefont {Colin}, \citenamefont {Gemmell}, \citenamefont {Lucas},
  \citenamefont {Leftwich},\  and\ \citenamefont
  {Costello}}]{Dabiri_etal_2019}%
  \BibitemOpen
  \bibfield  {author} {\bibinfo {author} {\bibfnamefont {J.~O.}\ \bibnamefont
  {Dabiri}}, \bibinfo {author} {\bibfnamefont {S.~P.}\ \bibnamefont {Colin}},
  \bibinfo {author} {\bibfnamefont {B.~J.}\ \bibnamefont {Gemmell}}, \bibinfo
  {author} {\bibfnamefont {K.~N.}\ \bibnamefont {Lucas}}, \bibinfo {author}
  {\bibfnamefont {M.~C.}\ \bibnamefont {Leftwich}},  and\ \bibinfo {author}
  {\bibfnamefont {J.~H.}\ \bibnamefont {Costello}},\ }\bibfield  {title}
  {\bibinfo {title} {Primitive and modern swimmers solve the challenges of
  turning similarly to achieve high manoeuverability},\ }\href@noop {}
  {\bibfield  {journal} {\bibinfo  {journal} {submitted}\ } (\bibinfo {year}
  {2019})}\BibitemShut {NoStop}%
\bibitem [{\citenamefont {Ren}\ \emph {et~al.}(2016)\citenamefont {Ren},
  \citenamefont {Dong}, \citenamefont {Deng},\  and\ \citenamefont
  {Tobalske}}]{Ren_etal_2016}%
  \BibitemOpen
  \bibfield  {author} {\bibinfo {author} {\bibfnamefont {Y.}~\bibnamefont
  {Ren}}, \bibinfo {author} {\bibfnamefont {H.}~\bibnamefont {Dong}}, \bibinfo
  {author} {\bibfnamefont {X.}~\bibnamefont {Deng}},  and\ \bibinfo {author}
  {\bibfnamefont {B.}~\bibnamefont {Tobalske}},\ }\bibfield  {title} {\bibinfo
  {title} {Turning on a dime: asymmetric vortex formation in hummingbird
  maneuvering flight},\ }\href@noop {} {\bibfield  {journal} {\bibinfo
  {journal} {Physical Review Fluids}\ }\textbf {\bibinfo {volume} {1}},\
  \bibinfo {pages} {050511} (\bibinfo {year} {2016})}\BibitemShut {NoStop}%
\bibitem [{\citenamefont {Eldredge}(2019)}]{Eldredge2019}%
  \BibitemOpen
  \bibfield  {author} {\bibinfo {author} {\bibfnamefont {J.~D.}\ \bibnamefont
  {Eldredge}},\ }\href@noop {} {\emph {\bibinfo {title} {Mathematical Modeling
  of Unsteady Inviscid Flows}}}\ (\bibinfo  {publisher} {Springer
  International},\ \bibinfo {year} {2019})\BibitemShut {NoStop}%
\bibitem [{\citenamefont {Mittal}\ \emph {et~al.}(2013)\citenamefont {Mittal},
  \citenamefont {Erath},\  and\ \citenamefont {Plesniak}}]{Mittal_etal_2013}%
  \BibitemOpen
  \bibfield  {author} {\bibinfo {author} {\bibfnamefont {R.}~\bibnamefont
  {Mittal}}, \bibinfo {author} {\bibfnamefont {B.~D.}\ \bibnamefont {Erath}},
  and\ \bibinfo {author} {\bibfnamefont {M.~W.}\ \bibnamefont {Plesniak}},\
  }\bibfield  {title} {\bibinfo {title} {Fluid dynamics of human phonation and
  speech},\ }\href@noop {} {\bibfield  {journal} {\bibinfo  {journal} {Annual
  Review of Fluid Mechanics}\ }\textbf {\bibinfo {volume} {45}},\ \bibinfo
  {pages} {437} (\bibinfo {year} {2013})}\BibitemShut {NoStop}%
\bibitem [{\citenamefont {Obrist}(2019)}]{Obrist2019}%
  \BibitemOpen
  \bibfield  {author} {\bibinfo {author} {\bibfnamefont {D.}~\bibnamefont
  {Obrist}},\ }\bibfield  {title} {\bibinfo {title} {Flow phenomena in the
  inner ear},\ }\href@noop {} {\bibfield  {journal} {\bibinfo  {journal}
  {Annual Review of Fluid Mechanics}\ }\textbf {\bibinfo {volume} {51}},\
  \bibinfo {pages} {487} (\bibinfo {year} {2019})}\BibitemShut {NoStop}%
\bibitem [{\citenamefont {Koehl}\ \emph {et~al.}(2001)\citenamefont {Koehl},
  \citenamefont {Koseff}, \citenamefont {Crimaldi}, \citenamefont {McCay},
  \citenamefont {Cooper}, \citenamefont {Wiley},\  and\ \citenamefont
  {Moore}}]{Koehl_etal_2001}%
  \BibitemOpen
  \bibfield  {author} {\bibinfo {author} {\bibfnamefont {M.~A.~R.}\
  \bibnamefont {Koehl}}, \bibinfo {author} {\bibfnamefont {J.~R.}\ \bibnamefont
  {Koseff}}, \bibinfo {author} {\bibfnamefont {J.~P.}\ \bibnamefont
  {Crimaldi}}, \bibinfo {author} {\bibfnamefont {M.~G.}\ \bibnamefont {McCay}},
  \bibinfo {author} {\bibfnamefont {T.}~\bibnamefont {Cooper}}, \bibinfo
  {author} {\bibfnamefont {M.~B.}\ \bibnamefont {Wiley}},  and\ \bibinfo
  {author} {\bibfnamefont {P.~A.}\ \bibnamefont {Moore}},\ }\bibfield  {title}
  {\bibinfo {title} {Lobster sniffing: antennule design and hydrodynamic
  filtering of information in an odor plume},\ }\href@noop {} {\bibfield
  {journal} {\bibinfo  {journal} {Science}\ }\textbf {\bibinfo {volume}
  {294}},\ \bibinfo {pages} {1948} (\bibinfo {year} {2001})}\BibitemShut
  {NoStop}%
\bibitem [{\citenamefont {Mellon}(2007)}]{Mellon2007}%
  \BibitemOpen
  \bibfield  {author} {\bibinfo {author} {\bibfnamefont {D.}~\bibnamefont
  {Mellon}},\ }\bibfield  {title} {\bibinfo {title} {Combining dissimilar
  senses: central processing of hydrodynamic and chemosensory inputs in aquatic
  crustaceans},\ }\href@noop {} {\bibfield  {journal} {\bibinfo  {journal}
  {Biological Bulletin}\ }\textbf {\bibinfo {volume} {213}},\ \bibinfo {pages}
  {1} (\bibinfo {year} {2007})}\BibitemShut {NoStop}%
\bibitem [{\citenamefont {Page}\ \emph
  {et~al.}(2011{\natexlab{a}})\citenamefont {Page}, \citenamefont {Dickman},
  \citenamefont {Webster},\  and\ \citenamefont {Weissburg}}]{Page_etal_2011a}%
  \BibitemOpen
  \bibfield  {author} {\bibinfo {author} {\bibfnamefont {J.~L.}\ \bibnamefont
  {Page}}, \bibinfo {author} {\bibfnamefont {B.~D.}\ \bibnamefont {Dickman}},
  \bibinfo {author} {\bibfnamefont {D.~R.}\ \bibnamefont {Webster}},  and\
  \bibinfo {author} {\bibfnamefont {M.~J.}\ \bibnamefont {Weissburg}},\
  }\bibfield  {title} {\bibinfo {title} {Getting ahead: Context-dependent
  responses to odor filaments drives along-stream progress during odor tracking
  in blue crabs},\ }\href@noop {} {\bibfield  {journal} {\bibinfo  {journal}
  {Journal of Experimental Biology}\ }\textbf {\bibinfo {volume} {214}},\
  \bibinfo {pages} {1498} (\bibinfo {year} {2011}{\natexlab{a}})}\BibitemShut
  {NoStop}%
\bibitem [{\citenamefont {Page}\ \emph
  {et~al.}(2011{\natexlab{b}})\citenamefont {Page}, \citenamefont {Dickman},
  \citenamefont {Webster},\  and\ \citenamefont {Weissburg}}]{Page_etal_2011b}%
  \BibitemOpen
  \bibfield  {author} {\bibinfo {author} {\bibfnamefont {J.~L.}\ \bibnamefont
  {Page}}, \bibinfo {author} {\bibfnamefont {B.~D.}\ \bibnamefont {Dickman}},
  \bibinfo {author} {\bibfnamefont {D.~R.}\ \bibnamefont {Webster}},  and\
  \bibinfo {author} {\bibfnamefont {M.~J.}\ \bibnamefont {Weissburg}},\
  }\bibfield  {title} {\bibinfo {title} {Staying the course: Chemical signal
  spatial properties and concentration mediate cross-stream motion in turbulent
  plumes},\ }\href@noop {} {\bibfield  {journal} {\bibinfo  {journal} {Journal
  of Experimental Biology}\ }\textbf {\bibinfo {volume} {214}},\ \bibinfo
  {pages} {1513} (\bibinfo {year} {2011}{\natexlab{b}})}\BibitemShut {NoStop}%
\bibitem [{\citenamefont {Moore}\ \emph {et~al.}(1999)\citenamefont {Moore},
  \citenamefont {Fields},\  and\ \citenamefont {Yen}}]{Moore_etal_1999}%
  \BibitemOpen
  \bibfield  {author} {\bibinfo {author} {\bibfnamefont {P.~A.}\ \bibnamefont
  {Moore}}, \bibinfo {author} {\bibfnamefont {D.~M.}\ \bibnamefont {Fields}},
  and\ \bibinfo {author} {\bibfnamefont {J.}~\bibnamefont {Yen}},\ }\bibfield
  {title} {\bibinfo {title} {Physical constraints of chemoreception in foraging
  copepods},\ }\href@noop {} {\bibfield  {journal} {\bibinfo  {journal}
  {Limnology and Oceanography}\ }\textbf {\bibinfo {volume} {44}},\ \bibinfo
  {pages} {166} (\bibinfo {year} {1999})}\BibitemShut {NoStop}%
\bibitem [{\citenamefont {Webster}\  and\ \citenamefont
  {Weissburg}(2009)}]{WebsterWeissburg2009}%
  \BibitemOpen
  \bibfield  {author} {\bibinfo {author} {\bibfnamefont {D.~R.}\ \bibnamefont
  {Webster}} and\ \bibinfo {author} {\bibfnamefont {M.~J.}\ \bibnamefont
  {Weissburg}},\ }\bibfield  {title} {\bibinfo {title} {The hydrodynamics of
  chemical cues among aquatic organisms},\ }\href@noop {} {\bibfield  {journal}
  {\bibinfo  {journal} {Annual Review of Fluid Mechanics}\ }\textbf {\bibinfo
  {volume} {41}},\ \bibinfo {pages} {73} (\bibinfo {year} {2009})}\BibitemShut
  {NoStop}%
\bibitem [{\citenamefont {Ahmed}\ \emph {et~al.}(2010)\citenamefont {Ahmed},
  \citenamefont {Shimizu},\  and\ \citenamefont {Stocker}}]{Ahmed_etal_2010}%
  \BibitemOpen
  \bibfield  {author} {\bibinfo {author} {\bibfnamefont {T.}~\bibnamefont
  {Ahmed}}, \bibinfo {author} {\bibfnamefont {T.~S.}\ \bibnamefont {Shimizu}},
  and\ \bibinfo {author} {\bibfnamefont {R.}~\bibnamefont {Stocker}},\
  }\bibfield  {title} {\bibinfo {title} {Microfluidics for bacterial
  chemotaxis},\ }\href@noop {} {\bibfield  {journal} {\bibinfo  {journal}
  {Integrative Biology}\ }\textbf {\bibinfo {volume} {2}},\ \bibinfo {pages}
  {604} (\bibinfo {year} {2010})}\BibitemShut {NoStop}%
\bibitem [{\citenamefont {Pekkan}\ \emph {et~al.}(2016)\citenamefont {Pekkan},
  \citenamefont {Chang}, \citenamefont {Uslu}, \citenamefont {Mani},
  \citenamefont {Chen},\  and\ \citenamefont {Holzman}}]{Pekkan_etal_2016}%
  \BibitemOpen
  \bibfield  {author} {\bibinfo {author} {\bibfnamefont {K.}~\bibnamefont
  {Pekkan}}, \bibinfo {author} {\bibfnamefont {B.}~\bibnamefont {Chang}},
  \bibinfo {author} {\bibfnamefont {F.}~\bibnamefont {Uslu}}, \bibinfo {author}
  {\bibfnamefont {K.}~\bibnamefont {Mani}}, \bibinfo {author} {\bibfnamefont
  {C.~Y.}\ \bibnamefont {Chen}},  and\ \bibinfo {author} {\bibfnamefont
  {R.}~\bibnamefont {Holzman}},\ }\bibfield  {title} {\bibinfo {title}
  {Characterization of zebrafish larvae suction feeding flow using mu {PIV} and
  optical coherence tomography},\ }\href@noop {} {\bibfield  {journal}
  {\bibinfo  {journal} {Experiments in Fluids}\ }\textbf {\bibinfo {volume}
  {57}},\ \bibinfo {pages} {112} (\bibinfo {year} {2016})}\BibitemShut
  {NoStop}%
\bibitem [{\citenamefont {Xu}\ \emph {et~al.}(2018)\citenamefont {Xu},
  \citenamefont {Nielsen},\  and\ \citenamefont {Kiorboe}}]{Xu_etal_2018}%
  \BibitemOpen
  \bibfield  {author} {\bibinfo {author} {\bibfnamefont {J.~Y.}\ \bibnamefont
  {Xu}}, \bibinfo {author} {\bibfnamefont {L.~T.}\ \bibnamefont {Nielsen}},
  and\ \bibinfo {author} {\bibfnamefont {T.}~\bibnamefont {Kiorboe}},\
  }\bibfield  {title} {\bibinfo {title} {Foraging response and acclimation of
  ambush feeding and feeding-current feeding copepods to toxic
  dinoflagellates},\ }\href@noop {} {\bibfield  {journal} {\bibinfo  {journal}
  {Limnology and Oceanography}\ }\textbf {\bibinfo {volume} {63}},\ \bibinfo
  {pages} {1449} (\bibinfo {year} {2018})}\BibitemShut {NoStop}%
\bibitem [{\citenamefont {Brumley}\ \emph {et~al.}(2019)\citenamefont
  {Brumley}, \citenamefont {Carrara}, \citenamefont {Hein}, \citenamefont
  {Yawata}, \citenamefont {Levin},\  and\ \citenamefont
  {Stocker}}]{Brumley_etal_2019}%
  \BibitemOpen
  \bibfield  {author} {\bibinfo {author} {\bibfnamefont {D.~R.}\ \bibnamefont
  {Brumley}}, \bibinfo {author} {\bibfnamefont {F.}~\bibnamefont {Carrara}},
  \bibinfo {author} {\bibfnamefont {A.~M.}\ \bibnamefont {Hein}}, \bibinfo
  {author} {\bibfnamefont {Y.}~\bibnamefont {Yawata}}, \bibinfo {author}
  {\bibfnamefont {S.~A.}\ \bibnamefont {Levin}},  and\ \bibinfo {author}
  {\bibfnamefont {R.}~\bibnamefont {Stocker}},\ }\bibfield  {title} {\bibinfo
  {title} {Bacteria push the limits of chemotactic precision to navigate
  dynamic chemical gradients},\ }\href@noop {} {\bibfield  {journal} {\bibinfo
  {journal} {Proceedings of the National Academy of Sciences}\ }\textbf
  {\bibinfo {volume} {116}},\ \bibinfo {pages} {10792} (\bibinfo {year}
  {2019})}\BibitemShut {NoStop}%
\bibitem [{\citenamefont {Sanfilippo}\ \emph {et~al.}(2019)\citenamefont
  {Sanfilippo}, \citenamefont {Lorestani}, \citenamefont {Koch}, \citenamefont
  {Bratton}, \citenamefont {Siryaporn}, \citenamefont {Stone},\  and\
  \citenamefont {Gitai}}]{Sanfilippo_etal_2019}%
  \BibitemOpen
  \bibfield  {author} {\bibinfo {author} {\bibfnamefont {J.~E.}\ \bibnamefont
  {Sanfilippo}}, \bibinfo {author} {\bibfnamefont {A.}~\bibnamefont
  {Lorestani}}, \bibinfo {author} {\bibfnamefont {M.~D.}\ \bibnamefont {Koch}},
  \bibinfo {author} {\bibfnamefont {B.~P.}\ \bibnamefont {Bratton}}, \bibinfo
  {author} {\bibfnamefont {A.}~\bibnamefont {Siryaporn}}, \bibinfo {author}
  {\bibfnamefont {H.~A.}\ \bibnamefont {Stone}},  and\ \bibinfo {author}
  {\bibfnamefont {Z.}~\bibnamefont {Gitai}},\ }\bibfield  {title} {\bibinfo
  {title} {Microfluidic-based transcriptomics reveal force-independent
  bacterial rheosensing},\ }\href@noop {} {\bibfield  {journal} {\bibinfo
  {journal} {Nature Microbiology}\ ,\ \bibinfo {pages} {10.1038/s41564}}
  (\bibinfo {year} {2019})}\BibitemShut {NoStop}%
\bibitem [{\citenamefont {Triantafyllou}\ \emph {et~al.}(2016)\citenamefont
  {Triantafyllou}, \citenamefont {Weymouth},\  and\ \citenamefont
  {Miao}}]{Triantafyllou_etal_2016}%
  \BibitemOpen
  \bibfield  {author} {\bibinfo {author} {\bibfnamefont {M.~S.}\ \bibnamefont
  {Triantafyllou}}, \bibinfo {author} {\bibfnamefont {G.~D.}\ \bibnamefont
  {Weymouth}},  and\ \bibinfo {author} {\bibfnamefont {J.}~\bibnamefont
  {Miao}},\ }\bibfield  {title} {\bibinfo {title} {Biomimetic survival
  hydrodynamics and flow sensing},\ }\href@noop {} {\bibfield  {journal}
  {\bibinfo  {journal} {Annual Review of Fluid Mechanics}\ }\textbf {\bibinfo
  {volume} {48}},\ \bibinfo {pages} {1} (\bibinfo {year} {2016})}\BibitemShut
  {NoStop}%
\bibitem [{\citenamefont {Oteiza}\ \emph {et~al.}(2017)\citenamefont {Oteiza},
  \citenamefont {Odstrcil}, \citenamefont {Lauder}, \citenamefont {Portugues},\
   and\ \citenamefont {Engert}}]{Oteiza_etal_2017}%
  \BibitemOpen
  \bibfield  {author} {\bibinfo {author} {\bibfnamefont {P.}~\bibnamefont
  {Oteiza}}, \bibinfo {author} {\bibfnamefont {I.}~\bibnamefont {Odstrcil}},
  \bibinfo {author} {\bibfnamefont {G.~V.}\ \bibnamefont {Lauder}}, \bibinfo
  {author} {\bibfnamefont {R.}~\bibnamefont {Portugues}},  and\ \bibinfo
  {author} {\bibfnamefont {F.}~\bibnamefont {Engert}},\ }\bibfield  {title}
  {\bibinfo {title} {A novel mechanism for mechanosensory-based rheotaxis in
  larval zebrafish},\ }\href@noop {} {\bibfield  {journal} {\bibinfo  {journal}
  {Nature}\ }\textbf {\bibinfo {volume} {547}},\ \bibinfo {pages} {445}
  (\bibinfo {year} {2017})}\BibitemShut {NoStop}%
\bibitem [{\citenamefont {Krieg}\ \emph {et~al.}(2019)\citenamefont {Krieg},
  \citenamefont {Nelson},\  and\ \citenamefont {Mohseni}}]{Krieg_etal_2019}%
  \BibitemOpen
  \bibfield  {author} {\bibinfo {author} {\bibfnamefont {M.}~\bibnamefont
  {Krieg}}, \bibinfo {author} {\bibfnamefont {K.}~\bibnamefont {Nelson}},  and\
  \bibinfo {author} {\bibfnamefont {K.}~\bibnamefont {Mohseni}},\ }\bibfield
  {title} {\bibinfo {title} {Distributed sensing for fluid disturbance
  compensation and motion control of intelligent robots},\ }\href@noop {}
  {\bibfield  {journal} {\bibinfo  {journal} {Nature Machine Intelligence}\
  }\textbf {\bibinfo {volume} {1}},\ \bibinfo {pages} {10.1038/s42256}
  (\bibinfo {year} {2019})}\BibitemShut {NoStop}%
\bibitem [{\citenamefont {Mohren}\ \emph {et~al.}(2018)\citenamefont {Mohren},
  \citenamefont {Daniel}, \citenamefont {Brunton},\  and\ \citenamefont
  {Brunton}}]{Mohren_etal_2018}%
  \BibitemOpen
  \bibfield  {author} {\bibinfo {author} {\bibfnamefont {T.~L.}\ \bibnamefont
  {Mohren}}, \bibinfo {author} {\bibfnamefont {T.~L.}\ \bibnamefont {Daniel}},
  \bibinfo {author} {\bibfnamefont {S.~L.}\ \bibnamefont {Brunton}},  and\
  \bibinfo {author} {\bibfnamefont {B.~W.}\ \bibnamefont {Brunton}},\
  }\bibfield  {title} {\bibinfo {title} {Neural-inspired sensors enable sparse,
  efficient classification of spatiotemporal data},\ }\href@noop {} {\bibfield
  {journal} {\bibinfo  {journal} {Proceedings of the National Academy of
  Sciences}\ }\textbf {\bibinfo {volume} {115}},\ \bibinfo {pages} {10564}
  (\bibinfo {year} {2018})}\BibitemShut {NoStop}%
\bibitem [{\citenamefont {Colvert}\ \emph {et~al.}(2018)\citenamefont
  {Colvert}, \citenamefont {Alsalman},\  and\ \citenamefont
  {Kanso}}]{Colvert_etal_2018}%
  \BibitemOpen
  \bibfield  {author} {\bibinfo {author} {\bibfnamefont {B.}~\bibnamefont
  {Colvert}}, \bibinfo {author} {\bibfnamefont {M.}~\bibnamefont {Alsalman}},
  and\ \bibinfo {author} {\bibfnamefont {E.}~\bibnamefont {Kanso}},\ }\bibfield
   {title} {\bibinfo {title} {Classifying vortex wakes using neural networks},\
  }\href@noop {} {\bibfield  {journal} {\bibinfo  {journal} {Bioinspiration and
  Biomimetics}\ }\textbf {\bibinfo {volume} {13}},\ \bibinfo {pages} {025003}
  (\bibinfo {year} {2018})}\BibitemShut {NoStop}%
\bibitem [{\citenamefont {Purcell}(1977)}]{Purcell1977}%
  \BibitemOpen
  \bibfield  {author} {\bibinfo {author} {\bibfnamefont {E.~M.}\ \bibnamefont
  {Purcell}},\ }\bibfield  {title} {\bibinfo {title} {Life at low reynolds
  number},\ }\href@noop {} {\bibfield  {journal} {\bibinfo  {journal} {American
  Journal of Physics}\ }\textbf {\bibinfo {volume} {45}},\ \bibinfo {pages} {3}
  (\bibinfo {year} {1977})}\BibitemShut {NoStop}%
\bibitem [{May(2018)}]{Mayo2018}%
  \BibitemOpen
  \href@noop {} {\emph {\bibinfo {title} {Infertility}}},\ \bibinfo {type}
  {www.mayoclinic.org}\ (\bibinfo  {institution} {Mayo Clinic},\ \bibinfo
  {year} {2018})\BibitemShut {NoStop}%
\bibitem [{\citenamefont {Audu}\ \emph {et~al.}(2009)\citenamefont {Audu},
  \citenamefont {Massa}, \citenamefont {Bukar}, \citenamefont {El-Nafaty},\
  and\ \citenamefont {Sa'ad}}]{Audu_etal_2009}%
  \BibitemOpen
  \bibfield  {author} {\bibinfo {author} {\bibfnamefont {B.~M.}\ \bibnamefont
  {Audu}}, \bibinfo {author} {\bibfnamefont {A.~A.}\ \bibnamefont {Massa}},
  \bibinfo {author} {\bibfnamefont {M.}~\bibnamefont {Bukar}}, \bibinfo
  {author} {\bibfnamefont {A.~U.}\ \bibnamefont {El-Nafaty}},  and\ \bibinfo
  {author} {\bibfnamefont {S.~T.}\ \bibnamefont {Sa'ad}},\ }\bibfield  {title}
  {\bibinfo {title} {Prevalence of utero-tubal infertility},\ }\href@noop {}
  {\bibfield  {journal} {\bibinfo  {journal} {Journal of Obstetrics and
  Gynaecology}\ }\textbf {\bibinfo {volume} {29}},\ \bibinfo {pages} {326}
  (\bibinfo {year} {2009})}\BibitemShut {NoStop}%
\bibitem [{\citenamefont {Lipinski}\  and\ \citenamefont
  {Mohseni}(2009)}]{LipinksiMohseni2009}%
  \BibitemOpen
  \bibfield  {author} {\bibinfo {author} {\bibfnamefont {D.}~\bibnamefont
  {Lipinski}} and\ \bibinfo {author} {\bibfnamefont {K.}~\bibnamefont
  {Mohseni}},\ }\bibfield  {title} {\bibinfo {title} {Flow structures and fluid
  transport for the hydromedusae sarsia tubulosa and aequorea victoria},\
  }\href@noop {} {\bibfield  {journal} {\bibinfo  {journal} {Journal of
  Experimental Biology}\ }\textbf {\bibinfo {volume} {212}},\ \bibinfo {pages}
  {2436} (\bibinfo {year} {2009})}\BibitemShut {NoStop}%
\bibitem [{\citenamefont {Womersley}(1955)}]{Womersley1955}%
  \BibitemOpen
  \bibfield  {author} {\bibinfo {author} {\bibfnamefont {J.~R.}\ \bibnamefont
  {Womersley}},\ }\bibfield  {title} {\bibinfo {title} {Method for the
  calculation of velocity, rate of flow and viscous drag in arteries when the
  pressure gradient is known},\ }\href@noop {} {\bibfield  {journal} {\bibinfo
  {journal} {Journal of Physiology}\ }\textbf {\bibinfo {volume} {127}},\
  \bibinfo {pages} {553} (\bibinfo {year} {1955})}\BibitemShut {NoStop}%
\bibitem [{\citenamefont {Spagnolie}\  and\ \citenamefont
  {Lauga}(2012)}]{SpagnolieLauga_2012}%
  \BibitemOpen
  \bibfield  {author} {\bibinfo {author} {\bibfnamefont {S.~E.}\ \bibnamefont
  {Spagnolie}} and\ \bibinfo {author} {\bibfnamefont {E.}~\bibnamefont
  {Lauga}},\ }\bibfield  {title} {\bibinfo {title} {Hydrodynamics of
  self-propulsion near a boundary: predictions and accuracy of far-field
  approximations},\ }\href@noop {} {\bibfield  {journal} {\bibinfo  {journal}
  {Journal of Fluid Mechanics}\ }\textbf {\bibinfo {volume} {700}},\ \bibinfo
  {pages} {105} (\bibinfo {year} {2012})}\BibitemShut {NoStop}%
\bibitem [{\citenamefont {Li}\  and\ \citenamefont
  {Ardekani}(2014)}]{LiArdekani2014}%
  \BibitemOpen
  \bibfield  {author} {\bibinfo {author} {\bibfnamefont {G.~J.}\ \bibnamefont
  {Li}} and\ \bibinfo {author} {\bibfnamefont {A.~M.}\ \bibnamefont
  {Ardekani}},\ }\bibfield  {title} {\bibinfo {title} {Hydrodynamic interaction
  of microswimmers near a wall},\ }\href@noop {} {\bibfield  {journal}
  {\bibinfo  {journal} {Physical Review E}\ }\textbf {\bibinfo {volume} {90}},\
  \bibinfo {pages} {013010} (\bibinfo {year} {2014})}\BibitemShut {NoStop}%
\bibitem [{\citenamefont {Lauga}(2016)}]{Lauga2016}%
  \BibitemOpen
  \bibfield  {author} {\bibinfo {author} {\bibfnamefont {E.}~\bibnamefont
  {Lauga}},\ }\bibfield  {title} {\bibinfo {title} {Bacterial hydrodynamics},\
  }\href@noop {} {\bibfield  {journal} {\bibinfo  {journal} {Annual Review of
  Fluid Mechanics}\ }\textbf {\bibinfo {volume} {48}},\ \bibinfo {pages} {105}
  (\bibinfo {year} {2016})}\BibitemShut {NoStop}%
\bibitem [{\citenamefont {Qiu}\ \emph {et~al.}(2014)\citenamefont {Qiu},
  \citenamefont {Lee}, \citenamefont {Mark}, \citenamefont {Morozov},
  \citenamefont {Munster}, \citenamefont {Mierka}, \citenamefont {Turek},
  \citenamefont {Leshansky},\  and\ \citenamefont {Fischer}}]{Qiu_etal_2014}%
  \BibitemOpen
  \bibfield  {author} {\bibinfo {author} {\bibfnamefont {T.}~\bibnamefont
  {Qiu}}, \bibinfo {author} {\bibfnamefont {T.~C.}\ \bibnamefont {Lee}},
  \bibinfo {author} {\bibfnamefont {A.~G.}\ \bibnamefont {Mark}}, \bibinfo
  {author} {\bibfnamefont {K.~I.}\ \bibnamefont {Morozov}}, \bibinfo {author}
  {\bibfnamefont {R.}~\bibnamefont {Munster}}, \bibinfo {author} {\bibfnamefont
  {O.}~\bibnamefont {Mierka}}, \bibinfo {author} {\bibfnamefont
  {S.}~\bibnamefont {Turek}}, \bibinfo {author} {\bibfnamefont {A.~M.}\
  \bibnamefont {Leshansky}},  and\ \bibinfo {author} {\bibfnamefont
  {P.}~\bibnamefont {Fischer}},\ }\bibfield  {title} {\bibinfo {title}
  {Swimming by reciprocal motion at low reynolds number},\ }\href@noop {}
  {\bibfield  {journal} {\bibinfo  {journal} {Nature Communications}\ }\textbf
  {\bibinfo {volume} {5}},\ \bibinfo {pages} {5119} (\bibinfo {year}
  {2014})}\BibitemShut {NoStop}%
\bibitem [{\citenamefont {Yu}\ \emph {et~al.}(2006)\citenamefont {Yu},
  \citenamefont {Lauga},\  and\ \citenamefont {Hosoi}}]{Yu_etal_2006}%
  \BibitemOpen
  \bibfield  {author} {\bibinfo {author} {\bibfnamefont {T.~S.}\ \bibnamefont
  {Yu}}, \bibinfo {author} {\bibfnamefont {E.}~\bibnamefont {Lauga}},  and\
  \bibinfo {author} {\bibfnamefont {A.~E.}\ \bibnamefont {Hosoi}},\ }\bibfield
  {title} {\bibinfo {title} {Experimental investigations of elastic tail
  propulsion at low reynolds number},\ }\href@noop {} {\bibfield  {journal}
  {\bibinfo  {journal} {Physics of Fluids}\ }\textbf {\bibinfo {volume} {18}},\
  \bibinfo {pages} {091701} (\bibinfo {year} {2006})}\BibitemShut {NoStop}%
\bibitem [{\citenamefont {Sznitman}\ \emph {et~al.}(2010)\citenamefont
  {Sznitman}, \citenamefont {Purohit}, \citenamefont {Krajacic}, \citenamefont
  {Lamitina},\  and\ \citenamefont {Arratia}}]{Sznitman_etal_2010}%
  \BibitemOpen
  \bibfield  {author} {\bibinfo {author} {\bibfnamefont {J.}~\bibnamefont
  {Sznitman}}, \bibinfo {author} {\bibfnamefont {P.~K.}\ \bibnamefont
  {Purohit}}, \bibinfo {author} {\bibfnamefont {P.}~\bibnamefont {Krajacic}},
  \bibinfo {author} {\bibfnamefont {T.}~\bibnamefont {Lamitina}},  and\
  \bibinfo {author} {\bibfnamefont {P.~E.}\ \bibnamefont {Arratia}},\
  }\bibfield  {title} {\bibinfo {title} {Material properties of caenorhabditis
  elegans swimming at low reynolds number},\ }\href@noop {} {\bibfield
  {journal} {\bibinfo  {journal} {Biophysical Journal}\ }\textbf {\bibinfo
  {volume} {98}},\ \bibinfo {pages} {617} (\bibinfo {year} {2010})}\BibitemShut
  {NoStop}%
\bibitem [{\citenamefont {et~al.}(2016)}]{Zhou_etal_2016}%
  \BibitemOpen
  \bibfield  {author} {\bibinfo {author} {\bibfnamefont {Q.~Z.}\ \bibnamefont
  {et~al.}},\ }\bibfield  {title} {\bibinfo {title} {Complete meiosis from
  embryonic stem cell-derived germ cells in vitro},\ }\href@noop {} {\bibfield
  {journal} {\bibinfo  {journal} {Cell Stem Cell}\ }\textbf {\bibinfo {volume}
  {18}},\ \bibinfo {pages} {330} (\bibinfo {year} {2016})}\BibitemShut
  {NoStop}%
\bibitem [{\citenamefont {Dreyfus}\ \emph {et~al.}(2005)\citenamefont
  {Dreyfus}, \citenamefont {Baudry}, \citenamefont {Roper}, \citenamefont
  {Fermigier}, \citenamefont {Stone},\  and\ \citenamefont
  {Bibette}}]{Dreyfus_etal_2005}%
  \BibitemOpen
  \bibfield  {author} {\bibinfo {author} {\bibfnamefont {R.}~\bibnamefont
  {Dreyfus}}, \bibinfo {author} {\bibfnamefont {J.}~\bibnamefont {Baudry}},
  \bibinfo {author} {\bibfnamefont {M.~L.}\ \bibnamefont {Roper}}, \bibinfo
  {author} {\bibfnamefont {M.}~\bibnamefont {Fermigier}}, \bibinfo {author}
  {\bibfnamefont {H.~A.}\ \bibnamefont {Stone}},  and\ \bibinfo {author}
  {\bibfnamefont {J.}~\bibnamefont {Bibette}},\ }\bibfield  {title} {\bibinfo
  {title} {Microscopic artificial swimmers},\ }\href@noop {} {\bibfield
  {journal} {\bibinfo  {journal} {Nature}\ }\textbf {\bibinfo {volume} {437}},\
  \bibinfo {pages} {862} (\bibinfo {year} {2005})}\BibitemShut {NoStop}%
\bibitem [{\citenamefont {Hove}\ \emph {et~al.}(2003)\citenamefont {Hove},
  \citenamefont {Koster}, \citenamefont {Forouhar}, \citenamefont
  {Acevedo-Bolton}, \citenamefont {Fraser},\  and\ \citenamefont
  {Gharib}}]{Hove_etal_2003}%
  \BibitemOpen
  \bibfield  {author} {\bibinfo {author} {\bibfnamefont {J.~R.}\ \bibnamefont
  {Hove}}, \bibinfo {author} {\bibfnamefont {R.~W.}\ \bibnamefont {Koster}},
  \bibinfo {author} {\bibfnamefont {A.~S.}\ \bibnamefont {Forouhar}}, \bibinfo
  {author} {\bibfnamefont {G.}~\bibnamefont {Acevedo-Bolton}}, \bibinfo
  {author} {\bibfnamefont {S.~E.}\ \bibnamefont {Fraser}},  and\ \bibinfo
  {author} {\bibfnamefont {M.}~\bibnamefont {Gharib}},\ }\bibfield  {title}
  {\bibinfo {title} {Intracardiac fluid forces are an essential epigenetic
  factor for embryonic cardiogenesis},\ }\href@noop {} {\bibfield  {journal}
  {\bibinfo  {journal} {Nature}\ }\textbf {\bibinfo {volume} {421}},\ \bibinfo
  {pages} {172} (\bibinfo {year} {2003})}\BibitemShut {NoStop}%
\bibitem [{\citenamefont {Wolpert}\  and\ \citenamefont
  {Tickle}(2010)}]{WolpertTickle2010}%
  \BibitemOpen
  \bibfield  {author} {\bibinfo {author} {\bibfnamefont {L.}~\bibnamefont
  {Wolpert}} and\ \bibinfo {author} {\bibfnamefont {C.}~\bibnamefont
  {Tickle}},\ }\href@noop {} {\emph {\bibinfo {title} {Principles of
  Development}}}\ (\bibinfo  {publisher} {Oxford University Press},\ \bibinfo
  {year} {2010})\BibitemShut {NoStop}%
\bibitem [{\citenamefont {Fung}(1993)}]{Fung1993}%
  \BibitemOpen
  \bibfield  {author} {\bibinfo {author} {\bibfnamefont {Y.~C.}\ \bibnamefont
  {Fung}},\ }\href@noop {} {\emph {\bibinfo {title} {Biomechanics: Mechanical
  Properties of Living Tissues}}}\ (\bibinfo  {publisher} {Springer-Verlag, New
  York},\ \bibinfo {year} {1993})\BibitemShut {NoStop}%
\bibitem [{\citenamefont {Liu}\  and\ \citenamefont
  {Katz}(2006)}]{LiuKatz2006}%
  \BibitemOpen
  \bibfield  {author} {\bibinfo {author} {\bibfnamefont {X.}~\bibnamefont
  {Liu}} and\ \bibinfo {author} {\bibfnamefont {J.}~\bibnamefont {Katz}},\
  }\bibfield  {title} {\bibinfo {title} {Instantaneous pressure and material
  acceleration measurements using a four-exposure {PIV} system},\ }\href@noop
  {} {\bibfield  {journal} {\bibinfo  {journal} {Experiments in Fluids}\
  }\textbf {\bibinfo {volume} {41}},\ \bibinfo {pages} {227} (\bibinfo {year}
  {2006})}\BibitemShut {NoStop}%
\bibitem [{\citenamefont {van Oudheusden}(2013)}]{vanOudheusden2013}%
  \BibitemOpen
  \bibfield  {author} {\bibinfo {author} {\bibfnamefont {B.}~\bibnamefont {van
  Oudheusden}},\ }\bibfield  {title} {\bibinfo {title} {{PIV}-based pressure
  measurement},\ }\href@noop {} {\bibfield  {journal} {\bibinfo  {journal}
  {Measurement Science and Technology}\ }\textbf {\bibinfo {volume} {24}},\
  \bibinfo {pages} {032001} (\bibinfo {year} {2013})}\BibitemShut {NoStop}%
\bibitem [{\citenamefont {Briggs}\ \emph {et~al.}(2018)\citenamefont {Briggs},
  \citenamefont {Weinreb}, \citenamefont {Wagner}, \citenamefont {Megason},
  \citenamefont {Peshkin}, \citenamefont {Kirschner},\  and\ \citenamefont
  {Klein}}]{Briggs_etal_2018}%
  \BibitemOpen
  \bibfield  {author} {\bibinfo {author} {\bibfnamefont {J.~A.}\ \bibnamefont
  {Briggs}}, \bibinfo {author} {\bibfnamefont {C.}~\bibnamefont {Weinreb}},
  \bibinfo {author} {\bibfnamefont {D.~E.}\ \bibnamefont {Wagner}}, \bibinfo
  {author} {\bibfnamefont {S.}~\bibnamefont {Megason}}, \bibinfo {author}
  {\bibfnamefont {L.}~\bibnamefont {Peshkin}}, \bibinfo {author} {\bibfnamefont
  {M.~W.}\ \bibnamefont {Kirschner}},  and\ \bibinfo {author} {\bibfnamefont
  {A.~M.}\ \bibnamefont {Klein}},\ }\bibfield  {title} {\bibinfo {title} {The
  dynamics of gene expression in vertebrate embryogenesis at single-cell
  resolution},\ }\href@noop {} {\bibfield  {journal} {\bibinfo  {journal}
  {Science}\ }\textbf {\bibinfo {volume} {360}},\ \bibinfo {pages} {eaar5780}
  (\bibinfo {year} {2018})}\BibitemShut {NoStop}%
\bibitem [{\citenamefont {Cieri}\  and\ \citenamefont
  {Farmer}(2016)}]{CieriFarmer2016}%
  \BibitemOpen
  \bibfield  {author} {\bibinfo {author} {\bibfnamefont {R.~L.}\ \bibnamefont
  {Cieri}} and\ \bibinfo {author} {\bibfnamefont {C.~G.}\ \bibnamefont
  {Farmer}},\ }\bibfield  {title} {\bibinfo {title} {Unidirectional pulmonary
  airflow in vertebrates: a review of structure, function, and evolution},\
  }\href@noop {} {\bibfield  {journal} {\bibinfo  {journal} {Journal of
  Comparative Physiology B}\ }\textbf {\bibinfo {volume} {186}},\ \bibinfo
  {pages} {541} (\bibinfo {year} {2016})}\BibitemShut {NoStop}%
\bibitem [{\citenamefont {Quinney}(1997)}]{Quinney1997}%
  \BibitemOpen
  \bibfield  {author} {\bibinfo {author} {\bibfnamefont {D.~A.}\ \bibnamefont
  {Quinney}},\ }\href@noop {} {\emph {\bibinfo {title} {Daniel Bernoulli and
  the making of the fluid equation}}},\ \bibinfo {type} {Tech. Rep.}\ (\bibinfo
   {institution} {Plus Magazine},\ \bibinfo {year} {1997})\BibitemShut
  {NoStop}%
\bibitem [{\citenamefont {Grotberg}\  and\ \citenamefont
  {Jensen}(2004)}]{GrotbergJensen2004}%
  \BibitemOpen
  \bibfield  {author} {\bibinfo {author} {\bibfnamefont {J.~B.}\ \bibnamefont
  {Grotberg}} and\ \bibinfo {author} {\bibfnamefont {O.~E.}\ \bibnamefont
  {Jensen}},\ }\bibfield  {title} {\bibinfo {title} {Biofluid mechanics in
  flexible tubes},\ }\href@noop {} {\bibfield  {journal} {\bibinfo  {journal}
  {Annual Review of Fluid Mechanics}\ }\textbf {\bibinfo {volume} {36}},\
  \bibinfo {pages} {121} (\bibinfo {year} {2004})}\BibitemShut {NoStop}%
\bibitem [{\citenamefont {Marsden}(2014)}]{Marsden2014}%
  \BibitemOpen
  \bibfield  {author} {\bibinfo {author} {\bibfnamefont {A.~L.}\ \bibnamefont
  {Marsden}},\ }\bibfield  {title} {\bibinfo {title} {Optimization in
  cardiovascular modeling},\ }\href@noop {} {\bibfield  {journal} {\bibinfo
  {journal} {Annual Review of Fluid Mechanics}\ }\textbf {\bibinfo {volume}
  {46}},\ \bibinfo {pages} {519} (\bibinfo {year} {2014})}\BibitemShut
  {NoStop}%
\bibitem [{\citenamefont {Mestre}\ \emph {et~al.}(2018)\citenamefont {Mestre},
  \citenamefont {Tithof}, \citenamefont {Du}, \citenamefont {Song},
  \citenamefont {Peng}, \citenamefont {Sweeney}, \citenamefont {Olveda},
  \citenamefont {Thomas}, \citenamefont {Nedergaard},\  and\ \citenamefont
  {Kelley}}]{Mestre_etal_2018}%
  \BibitemOpen
  \bibfield  {author} {\bibinfo {author} {\bibfnamefont {H.}~\bibnamefont
  {Mestre}}, \bibinfo {author} {\bibfnamefont {J.}~\bibnamefont {Tithof}},
  \bibinfo {author} {\bibfnamefont {T.}~\bibnamefont {Du}}, \bibinfo {author}
  {\bibfnamefont {W.}~\bibnamefont {Song}}, \bibinfo {author} {\bibfnamefont
  {W.}~\bibnamefont {Peng}}, \bibinfo {author} {\bibfnamefont {A.~M.}\
  \bibnamefont {Sweeney}}, \bibinfo {author} {\bibfnamefont {G.}~\bibnamefont
  {Olveda}}, \bibinfo {author} {\bibfnamefont {J.~H.}\ \bibnamefont {Thomas}},
  \bibinfo {author} {\bibfnamefont {M.}~\bibnamefont {Nedergaard}},  and\
  \bibinfo {author} {\bibfnamefont {D.~H.}\ \bibnamefont {Kelley}},\ }\bibfield
   {title} {\bibinfo {title} {Flow of cerebrospinal fluid is driven by arterial
  pulsations and is reduced in hypertension},\ }\href@noop {} {\bibfield
  {journal} {\bibinfo  {journal} {Nature Communications}\ }\textbf {\bibinfo
  {volume} {9}},\ \bibinfo {pages} {4878} (\bibinfo {year} {2018})}\BibitemShut
  {NoStop}%
\bibitem [{\citenamefont {Staples}\  and\ \citenamefont
  {Mikel-Stites}(2018)}]{StaplesMikel_2018}%
  \BibitemOpen
  \bibfield  {author} {\bibinfo {author} {\bibfnamefont {A.}~\bibnamefont
  {Staples}} and\ \bibinfo {author} {\bibfnamefont {M.}~\bibnamefont
  {Mikel-Stites}},\ }\bibfield  {title} {\bibinfo {title} {Ant-man and the
  wasp: Microscale respiration and microfluidic technology},\ }\href@noop {}
  {\bibfield  {journal} {\bibinfo  {journal} {Superhero Science and
  Technology}\ }\textbf {\bibinfo {volume} {1}},\ \bibinfo {pages} {1}
  (\bibinfo {year} {2018})}\BibitemShut {NoStop}%
\bibitem [{\citenamefont {Stocker}(2012)}]{Stocker2012}%
  \BibitemOpen
  \bibfield  {author} {\bibinfo {author} {\bibfnamefont {R.}~\bibnamefont
  {Stocker}},\ }\bibfield  {title} {\bibinfo {title} {Marine microbes see a sea
  of gradients},\ }\href@noop {} {\bibfield  {journal} {\bibinfo  {journal}
  {Science}\ }\textbf {\bibinfo {volume} {338}},\ \bibinfo {pages} {628}
  (\bibinfo {year} {2012})}\BibitemShut {NoStop}%
\bibitem [{\citenamefont {Nepf}(2012)}]{Nepf2012}%
  \BibitemOpen
  \bibfield  {author} {\bibinfo {author} {\bibfnamefont {H.~M.}\ \bibnamefont
  {Nepf}},\ }\bibfield  {title} {\bibinfo {title} {Flow and transport in
  regions with aquatic vegetation},\ }\href@noop {} {\bibfield  {journal}
  {\bibinfo  {journal} {Annual Review of Fluid Mechanics}\ }\textbf {\bibinfo
  {volume} {44}},\ \bibinfo {pages} {123} (\bibinfo {year} {2012})}\BibitemShut
  {NoStop}%
\bibitem [{\citenamefont {Sinhuber}\  and\ \citenamefont
  {Ouellette}(2017)}]{SinhuberOuellette2017}%
  \BibitemOpen
  \bibfield  {author} {\bibinfo {author} {\bibfnamefont {M.}~\bibnamefont
  {Sinhuber}} and\ \bibinfo {author} {\bibfnamefont {N.~T.}\ \bibnamefont
  {Ouellette}},\ }\bibfield  {title} {\bibinfo {title} {Phase coexistence in
  insect swarms},\ }\href@noop {} {\bibfield  {journal} {\bibinfo  {journal}
  {Physical Review Letters}\ }\textbf {\bibinfo {volume} {119}},\ \bibinfo
  {pages} {178003} (\bibinfo {year} {2017})}\BibitemShut {NoStop}%
\bibitem [{\citenamefont {Houghton}\ \emph {et~al.}(2018)\citenamefont
  {Houghton}, \citenamefont {Koseff}, \citenamefont {Monismith},\  and\
  \citenamefont {Dabiri}}]{Houghton_etal_2018}%
  \BibitemOpen
  \bibfield  {author} {\bibinfo {author} {\bibfnamefont {I.~A.}\ \bibnamefont
  {Houghton}}, \bibinfo {author} {\bibfnamefont {J.~R.}\ \bibnamefont
  {Koseff}}, \bibinfo {author} {\bibfnamefont {S.~G.}\ \bibnamefont
  {Monismith}},  and\ \bibinfo {author} {\bibfnamefont {J.~O.}\ \bibnamefont
  {Dabiri}},\ }\bibfield  {title} {\bibinfo {title} {Vertically migrating
  swimmers generate aggregation-scale eddies in a stratified column},\
  }\href@noop {} {\bibfield  {journal} {\bibinfo  {journal} {Nature}\ }\textbf
  {\bibinfo {volume} {556}},\ \bibinfo {pages} {497} (\bibinfo {year}
  {2018})}\BibitemShut {NoStop}%
\bibitem [{\citenamefont {Shapiro}\ \emph {et~al.}(2014)\citenamefont
  {Shapiro}, \citenamefont {Fernandez}, \citenamefont {Garren}, \citenamefont
  {Guasto}, \citenamefont {Debaillon-Vesque}, \citenamefont {Kramarsky-Winter},
  \citenamefont {Vardi},\  and\ \citenamefont {Stocker}}]{Shapiro_etal_2014}%
  \BibitemOpen
  \bibfield  {author} {\bibinfo {author} {\bibfnamefont {O.~H.}\ \bibnamefont
  {Shapiro}}, \bibinfo {author} {\bibfnamefont {V.~I.}\ \bibnamefont
  {Fernandez}}, \bibinfo {author} {\bibfnamefont {M.}~\bibnamefont {Garren}},
  \bibinfo {author} {\bibfnamefont {J.~S.}\ \bibnamefont {Guasto}}, \bibinfo
  {author} {\bibfnamefont {F.~P.}\ \bibnamefont {Debaillon-Vesque}}, \bibinfo
  {author} {\bibfnamefont {E.}~\bibnamefont {Kramarsky-Winter}}, \bibinfo
  {author} {\bibfnamefont {A.}~\bibnamefont {Vardi}},  and\ \bibinfo {author}
  {\bibfnamefont {R.}~\bibnamefont {Stocker}},\ }\bibfield  {title} {\bibinfo
  {title} {Vortical ciliary flows actively enhance mass transport in reef
  corals},\ }\href@noop {} {\bibfield  {journal} {\bibinfo  {journal}
  {Proceedings of the National Academy of Sciences}\ }\textbf {\bibinfo
  {volume} {111}},\ \bibinfo {pages} {13391} (\bibinfo {year}
  {2014})}\BibitemShut {NoStop}%
\bibitem [{\citenamefont {Katija}(2012)}]{Katija2012}%
  \BibitemOpen
  \bibfield  {author} {\bibinfo {author} {\bibfnamefont {K.}~\bibnamefont
  {Katija}},\ }\bibfield  {title} {\bibinfo {title} {Biogenic inputs to ocean
  mixing},\ }\href@noop {} {\bibfield  {journal} {\bibinfo  {journal} {Journal
  of Experimental Biology}\ }\textbf {\bibinfo {volume} {215}},\ \bibinfo
  {pages} {1040} (\bibinfo {year} {2012})}\BibitemShut {NoStop}%
\bibitem [{\citenamefont {Filella}\ \emph {et~al.}(2018)\citenamefont
  {Filella}, \citenamefont {Nadal}, \citenamefont {Sire}, \citenamefont
  {Kanso},\  and\ \citenamefont {Eloy}}]{Filella_etal_2018}%
  \BibitemOpen
  \bibfield  {author} {\bibinfo {author} {\bibfnamefont {A.}~\bibnamefont
  {Filella}}, \bibinfo {author} {\bibfnamefont {F.}~\bibnamefont {Nadal}},
  \bibinfo {author} {\bibfnamefont {C.}~\bibnamefont {Sire}}, \bibinfo {author}
  {\bibfnamefont {E.}~\bibnamefont {Kanso}},  and\ \bibinfo {author}
  {\bibfnamefont {C.}~\bibnamefont {Eloy}},\ }\bibfield  {title} {\bibinfo
  {title} {Model of collective fish behavior with hydrodynamic interactions},\
  }\href@noop {} {\bibfield  {journal} {\bibinfo  {journal} {Physical Review
  Letters}\ }\textbf {\bibinfo {volume} {120}},\ \bibinfo {pages} {198101}
  (\bibinfo {year} {2018})}\BibitemShut {NoStop}%
\bibitem [{\citenamefont {Morrell}\ \emph {et~al.}(2019)\citenamefont
  {Morrell}, \citenamefont {Spagnolie},\  and\ \citenamefont
  {Thiffeault}}]{Morrell_etal_2019}%
  \BibitemOpen
  \bibfield  {author} {\bibinfo {author} {\bibfnamefont {T.~A.}\ \bibnamefont
  {Morrell}}, \bibinfo {author} {\bibfnamefont {S.~E.}\ \bibnamefont
  {Spagnolie}},  and\ \bibinfo {author} {\bibfnamefont {J.~L.}\ \bibnamefont
  {Thiffeault}},\ }\bibfield  {title} {\bibinfo {title} {Velocity fluctuations
  in a dilute suspension of viscous vortex rings},\ }\href@noop {} {\bibfield
  {journal} {\bibinfo  {journal} {Physical Review Fluids}\ }\textbf {\bibinfo
  {volume} {4}},\ \bibinfo {pages} {044501} (\bibinfo {year}
  {2019})}\BibitemShut {NoStop}%
\bibitem [{\citenamefont {Wilhelmus}\ \emph {et~al.}(2019)\citenamefont
  {Wilhelmus}, \citenamefont {Nawroth}, \citenamefont {Rallabandi},\  and\
  \citenamefont {Dabiri}}]{Wilhelmus_etal_2019}%
  \BibitemOpen
  \bibfield  {author} {\bibinfo {author} {\bibfnamefont {M.~M.}\ \bibnamefont
  {Wilhelmus}}, \bibinfo {author} {\bibfnamefont {J.~C.}\ \bibnamefont
  {Nawroth}}, \bibinfo {author} {\bibfnamefont {B.}~\bibnamefont {Rallabandi}},
   and\ \bibinfo {author} {\bibfnamefont {J.~O.}\ \bibnamefont {Dabiri}},\
  }\bibfield  {title} {\bibinfo {title} {Effect of swarm configuration on fluid
  transport during vertical collective motion},\ }\href@noop {} {\bibfield
  {journal} {\bibinfo  {journal} {in revision}\ } (\bibinfo {year}
  {2019})}\BibitemShut {NoStop}%
\bibitem [{\citenamefont {Alben}\ \emph {et~al.}(2002)\citenamefont {Alben},
  \citenamefont {Shelley},\  and\ \citenamefont {Zhang}}]{Alben_etal_2002}%
  \BibitemOpen
  \bibfield  {author} {\bibinfo {author} {\bibfnamefont {S.}~\bibnamefont
  {Alben}}, \bibinfo {author} {\bibfnamefont {M.}~\bibnamefont {Shelley}},
  and\ \bibinfo {author} {\bibfnamefont {J.}~\bibnamefont {Zhang}},\ }\bibfield
   {title} {\bibinfo {title} {Drag reduction through self-similar bending of a
  flexible body},\ }\href@noop {} {\bibfield  {journal} {\bibinfo  {journal}
  {Nature}\ }\textbf {\bibinfo {volume} {420}},\ \bibinfo {pages} {479}
  (\bibinfo {year} {2002})}\BibitemShut {NoStop}%
\bibitem [{\citenamefont {et~al.}(2009)}]{Barthelmie_etal_2009}%
  \BibitemOpen
  \bibfield  {author} {\bibinfo {author} {\bibfnamefont {R.~J.~B.}\
  \bibnamefont {et~al.}},\ }\bibfield  {title} {\bibinfo {title} {Modelling and
  measuring flow and wind turbine wakes in large wind farms offshore},\
  }\href@noop {} {\bibfield  {journal} {\bibinfo  {journal} {Wind Energy}\
  }\textbf {\bibinfo {volume} {12}},\ \bibinfo {pages} {431} (\bibinfo {year}
  {2009})}\BibitemShut {NoStop}%
\bibitem [{\citenamefont {Howland}\ \emph {et~al.}(2019)\citenamefont
  {Howland}, \citenamefont {Lele},\  and\ \citenamefont
  {Dabiri}}]{Howland_etal_2019}%
  \BibitemOpen
  \bibfield  {author} {\bibinfo {author} {\bibfnamefont {M.~F.}\ \bibnamefont
  {Howland}}, \bibinfo {author} {\bibfnamefont {S.~K.}\ \bibnamefont {Lele}},
  and\ \bibinfo {author} {\bibfnamefont {J.~O.}\ \bibnamefont {Dabiri}},\
  }\bibfield  {title} {\bibinfo {title} {Wind farm power optimization through
  wake steering},\ }\href@noop {} {\bibfield  {journal} {\bibinfo  {journal}
  {Proceedings of the National Academy of Sciences}\ } (\bibinfo {year}
  {2019})}\BibitemShut {NoStop}%
\bibitem [{\citenamefont {Bush}\  and\ \citenamefont {Hu}(2006)}]{BushHu2006}%
  \BibitemOpen
  \bibfield  {author} {\bibinfo {author} {\bibfnamefont {J.~W.~M.}\
  \bibnamefont {Bush}} and\ \bibinfo {author} {\bibfnamefont {D.~L.}\
  \bibnamefont {Hu}},\ }\bibfield  {title} {\bibinfo {title} {Walking on water:
  Biolocomotion at the interface},\ }\href@noop {} {\bibfield  {journal}
  {\bibinfo  {journal} {Annual Review of Fluid Mechanics}\ }\textbf {\bibinfo
  {volume} {38}},\ \bibinfo {pages} {339} (\bibinfo {year} {2006})}\BibitemShut
  {NoStop}%
\bibitem [{\citenamefont {Hu}\ \emph {et~al.}(2010)\citenamefont {Hu},
  \citenamefont {Prakash}, \citenamefont {Chan},\  and\ \citenamefont
  {Bush}}]{Hu_etal_2010}%
  \BibitemOpen
  \bibfield  {author} {\bibinfo {author} {\bibfnamefont {D.~L.}\ \bibnamefont
  {Hu}}, \bibinfo {author} {\bibfnamefont {M.}~\bibnamefont {Prakash}},
  \bibinfo {author} {\bibfnamefont {B.}~\bibnamefont {Chan}},  and\ \bibinfo
  {author} {\bibfnamefont {J.~W.~M.}\ \bibnamefont {Bush}},\ }\bibinfo {title}
  {Water-walking devices},\ in\ \href@noop {} {\emph {\bibinfo {booktitle}
  {Animal Locomotion}}},\ \bibinfo {editor} {edited by\ \bibinfo {editor}
  {\bibfnamefont {G.~K.}\ \bibnamefont {Taylor}}, \bibinfo {editor}
  {\bibfnamefont {M.~S.}\ \bibnamefont {Triantafyllou}},  and\ \bibinfo
  {editor} {\bibfnamefont {C.}~\bibnamefont {Tropea}}}\ (\bibinfo  {publisher}
  {Springer-Verlag},\ \bibinfo {address} {Berlin},\ \bibinfo {year} {2010})\
  pp.\ \bibinfo {pages} {131--140}\BibitemShut {NoStop}%
\bibitem [{\citenamefont {Mendelson}\  and\ \citenamefont
  {Techet}(2018)}]{MendelsonTechet2018}%
  \BibitemOpen
  \bibfield  {author} {\bibinfo {author} {\bibfnamefont {L.}~\bibnamefont
  {Mendelson}} and\ \bibinfo {author} {\bibfnamefont {A.~H.}\ \bibnamefont
  {Techet}},\ }\bibfield  {title} {\bibinfo {title} {Multi-camera volumetric
  {PIV} for the study of jumping fish},\ }\href@noop {} {\bibfield  {journal}
  {\bibinfo  {journal} {Experiments in Fluids}\ }\textbf {\bibinfo {volume}
  {59}},\ \bibinfo {pages} {10} (\bibinfo {year} {2018})}\BibitemShut {NoStop}%
\bibitem [{\citenamefont {Krane}\  and\ \citenamefont
  {Wei}(2006)}]{KraneWei_2006}%
  \BibitemOpen
  \bibfield  {author} {\bibinfo {author} {\bibfnamefont {M.~H.}\ \bibnamefont
  {Krane}} and\ \bibinfo {author} {\bibfnamefont {T.}~\bibnamefont {Wei}},\
  }\bibfield  {title} {\bibinfo {title} {Theoretical assessment of unsteady
  aerodynamic effects in phonation},\ }\href@noop {} {\bibfield  {journal}
  {\bibinfo  {journal} {Journal of the Acoustical Society of America}\ }\textbf
  {\bibinfo {volume} {120}},\ \bibinfo {pages} {1578} (\bibinfo {year}
  {2006})}\BibitemShut {NoStop}%
\bibitem [{\citenamefont {Yaniv}\ \emph {et~al.}(2003)\citenamefont {Yaniv},
  \citenamefont {Elad}, \citenamefont {Jaffa},\  and\ \citenamefont
  {Eytan}}]{Yaniv_etal_2003}%
  \BibitemOpen
  \bibfield  {author} {\bibinfo {author} {\bibfnamefont {S.}~\bibnamefont
  {Yaniv}}, \bibinfo {author} {\bibfnamefont {D.}~\bibnamefont {Elad}},
  \bibinfo {author} {\bibfnamefont {A.~J.}\ \bibnamefont {Jaffa}},  and\
  \bibinfo {author} {\bibfnamefont {O.}~\bibnamefont {Eytan}},\ }\bibfield
  {title} {\bibinfo {title} {Biofluid aspects of embryo transfer},\ }\href@noop
  {} {\bibfield  {journal} {\bibinfo  {journal} {Annals of Biomedical
  Engineering}\ }\textbf {\bibinfo {volume} {31}},\ \bibinfo {pages} {1255}
  (\bibinfo {year} {2003})}\BibitemShut {NoStop}%
\bibitem [{\citenamefont {Lehn}\ \emph {et~al.}(2016)\citenamefont {Lehn},
  \citenamefont {Baumer},\  and\ \citenamefont {Leftwich}}]{Lehn_etal_2016}%
  \BibitemOpen
  \bibfield  {author} {\bibinfo {author} {\bibfnamefont {A.~M.}\ \bibnamefont
  {Lehn}}, \bibinfo {author} {\bibfnamefont {A.}~\bibnamefont {Baumer}},  and\
  \bibinfo {author} {\bibfnamefont {M.~C.}\ \bibnamefont {Leftwich}},\
  }\bibfield  {title} {\bibinfo {title} {An experimental approach to a
  simplified model of human birth},\ }\href@noop {} {\bibfield  {journal}
  {\bibinfo  {journal} {Journal of Biomechanics}\ }\textbf {\bibinfo {volume}
  {49}},\ \bibinfo {pages} {2313} (\bibinfo {year} {2016})}\BibitemShut
  {NoStop}%
\bibitem [{\citenamefont {Jiang}\ \emph {et~al.}(2018)\citenamefont {Jiang},
  \citenamefont {Nicolls}, \citenamefont {Tian},\  and\ \citenamefont
  {Rockson}}]{Jiang_etal_2018}%
  \BibitemOpen
  \bibfield  {author} {\bibinfo {author} {\bibfnamefont {X.~G.}\ \bibnamefont
  {Jiang}}, \bibinfo {author} {\bibfnamefont {M.~R.}\ \bibnamefont {Nicolls}},
  \bibinfo {author} {\bibfnamefont {W.}~\bibnamefont {Tian}},  and\ \bibinfo
  {author} {\bibfnamefont {S.~G.}\ \bibnamefont {Rockson}},\ }\bibfield
  {title} {\bibinfo {title} {Lymphatic dysfunction, leukotrienes, and
  lymphedema},\ }\href@noop {} {\bibfield  {journal} {\bibinfo  {journal}
  {Annual Review of Physiology}\ }\textbf {\bibinfo {volume} {80}},\ \bibinfo
  {pages} {49} (\bibinfo {year} {2018})}\BibitemShut {NoStop}%
\bibitem [{\citenamefont {Bar-On}\ \emph {et~al.}(2018)\citenamefont {Bar-On},
  \citenamefont {Phillips},\  and\ \citenamefont {Milo}}]{BarOn_etal_2018}%
  \BibitemOpen
  \bibfield  {author} {\bibinfo {author} {\bibfnamefont {Y.~M.}\ \bibnamefont
  {Bar-On}}, \bibinfo {author} {\bibfnamefont {R.}~\bibnamefont {Phillips}},
  and\ \bibinfo {author} {\bibfnamefont {R.}~\bibnamefont {Milo}},\ }\bibfield
  {title} {\bibinfo {title} {The biomass distribution on earth},\ }\href@noop
  {} {\bibfield  {journal} {\bibinfo  {journal} {Proceedings of the National
  Academy of Sciences}\ }\textbf {\bibinfo {volume} {115}},\ \bibinfo {pages}
  {6506} (\bibinfo {year} {2018})}\BibitemShut {NoStop}%
\bibitem [{\citenamefont {Pedley}(2010)}]{Pedley2010}%
  \BibitemOpen
  \bibfield  {author} {\bibinfo {author} {\bibfnamefont {T.~J.}\ \bibnamefont
  {Pedley}},\ }\bibfield  {title} {\bibinfo {title} {Collective behaviour of
  swimming micro-organisms},\ }\href@noop {} {\bibfield  {journal} {\bibinfo
  {journal} {Experimental Mechanics}\ }\textbf {\bibinfo {volume} {50}},\
  \bibinfo {pages} {1293} (\bibinfo {year} {2010})}\BibitemShut {NoStop}%
\bibitem [{\citenamefont {Nawroth}\ \emph {et~al.}(2017)\citenamefont
  {Nawroth}, \citenamefont {Guo}, \citenamefont {Koch}, \citenamefont
  {Heath-Heckman}, \citenamefont {Hermanson}, \citenamefont {Ruby},
  \citenamefont {Dabiri}, \citenamefont {Kanso},\  and\ \citenamefont
  {McFall-Ngai}}]{Nawroth_etal_2017}%
  \BibitemOpen
  \bibfield  {author} {\bibinfo {author} {\bibfnamefont {J.~C.}\ \bibnamefont
  {Nawroth}}, \bibinfo {author} {\bibfnamefont {H.}~\bibnamefont {Guo}},
  \bibinfo {author} {\bibfnamefont {E.}~\bibnamefont {Koch}}, \bibinfo {author}
  {\bibfnamefont {E.~A.}\ \bibnamefont {Heath-Heckman}}, \bibinfo {author}
  {\bibfnamefont {J.~C.}\ \bibnamefont {Hermanson}}, \bibinfo {author}
  {\bibfnamefont {E.}~\bibnamefont {Ruby}}, \bibinfo {author} {\bibfnamefont
  {J.~O.}\ \bibnamefont {Dabiri}}, \bibinfo {author} {\bibfnamefont
  {E.}~\bibnamefont {Kanso}},  and\ \bibinfo {author} {\bibfnamefont
  {M.}~\bibnamefont {McFall-Ngai}},\ }\bibfield  {title} {\bibinfo {title}
  {Motile cilia create fluid-mechanical microhabitats for the active
  recruitment of the host microbiome},\ }\href@noop {} {\bibfield  {journal}
  {\bibinfo  {journal} {Proceedings of the National Academy of Sciences}\
  }\textbf {\bibinfo {volume} {114}},\ \bibinfo {pages} {9510} (\bibinfo {year}
  {2017})}\BibitemShut {NoStop}%
\bibitem [{\citenamefont {Stamhuis}\  and\ \citenamefont
  {Videler}(1995)}]{StamhuisVideler1995}%
  \BibitemOpen
  \bibfield  {author} {\bibinfo {author} {\bibfnamefont {E.~J.}\ \bibnamefont
  {Stamhuis}} and\ \bibinfo {author} {\bibfnamefont {J.~J.}\ \bibnamefont
  {Videler}},\ }\bibfield  {title} {\bibinfo {title} {Quantitative
  flow-analysis around aquatic animals using laser sheet particle image
  velocimetry},\ }\href@noop {} {\bibfield  {journal} {\bibinfo  {journal}
  {Journal of Experimental Biology}\ }\textbf {\bibinfo {volume} {198}},\
  \bibinfo {pages} {283} (\bibinfo {year} {1995})}\BibitemShut {NoStop}%
\bibitem [{\citenamefont {Murphy}\ \emph {et~al.}(2012)\citenamefont {Murphy},
  \citenamefont {Webster},\  and\ \citenamefont {Yen}}]{Murphy_etal_2012}%
  \BibitemOpen
  \bibfield  {author} {\bibinfo {author} {\bibfnamefont {D.~W.}\ \bibnamefont
  {Murphy}}, \bibinfo {author} {\bibfnamefont {D.~R.}\ \bibnamefont {Webster}},
   and\ \bibinfo {author} {\bibfnamefont {J.}~\bibnamefont {Yen}},\ }\bibfield
  {title} {\bibinfo {title} {A high-speed tomographic piv system for measuring
  zooplanktonic flow},\ }\href@noop {} {\bibfield  {journal} {\bibinfo
  {journal} {Limnology and Oceanography: Methods}\ }\textbf {\bibinfo {volume}
  {10}},\ \bibinfo {pages} {1096–} (\bibinfo {year} {2012})}\BibitemShut
  {NoStop}%
\bibitem [{\citenamefont {Adhikari}\  and\ \citenamefont
  {Longmire}(2013)}]{AdhikariLongmore2013}%
  \BibitemOpen
  \bibfield  {author} {\bibinfo {author} {\bibfnamefont {D.}~\bibnamefont
  {Adhikari}} and\ \bibinfo {author} {\bibfnamefont {E.~K.}\ \bibnamefont
  {Longmire}},\ }\bibfield  {title} {\bibinfo {title} {Infrared tomographic
  {PIV} and {3D} motion tracking system applied to aquatic predator-prey
  interaction},\ }\href@noop {} {\bibfield  {journal} {\bibinfo  {journal}
  {Measurement Science and Technology}\ }\textbf {\bibinfo {volume} {24}},\
  \bibinfo {pages} {024011} (\bibinfo {year} {2013})}\BibitemShut {NoStop}%
\bibitem [{\citenamefont {Dabiri}\ \emph {et~al.}(2014)\citenamefont {Dabiri},
  \citenamefont {Bose}, \citenamefont {Gemmell}, \citenamefont {Colin},\  and\
  \citenamefont {Costello}}]{Dabiri_etal_2014}%
  \BibitemOpen
  \bibfield  {author} {\bibinfo {author} {\bibfnamefont {J.~O.}\ \bibnamefont
  {Dabiri}}, \bibinfo {author} {\bibfnamefont {S.}~\bibnamefont {Bose}},
  \bibinfo {author} {\bibfnamefont {B.~J.}\ \bibnamefont {Gemmell}}, \bibinfo
  {author} {\bibfnamefont {S.~P.}\ \bibnamefont {Colin}},  and\ \bibinfo
  {author} {\bibfnamefont {J.~H.}\ \bibnamefont {Costello}},\ }\bibfield
  {title} {\bibinfo {title} {An algorithm to estimate unsteady and quasi-steady
  pressure fields from velocity field measurements},\ }\href@noop {} {\bibfield
   {journal} {\bibinfo  {journal} {Journal of Experimental Biology}\ }\textbf
  {\bibinfo {volume} {217}},\ \bibinfo {pages} {331} (\bibinfo {year}
  {2014})}\BibitemShut {NoStop}%
\bibitem [{\citenamefont {Lucas}\ \emph {et~al.}(2017)\citenamefont {Lucas},
  \citenamefont {Dabiri},\  and\ \citenamefont {Lauder}}]{Lucas_etal_2017}%
  \BibitemOpen
  \bibfield  {author} {\bibinfo {author} {\bibfnamefont {K.~N.}\ \bibnamefont
  {Lucas}}, \bibinfo {author} {\bibfnamefont {J.~O.}\ \bibnamefont {Dabiri}},
  and\ \bibinfo {author} {\bibfnamefont {G.~V.}\ \bibnamefont {Lauder}},\
  }\bibfield  {title} {\bibinfo {title} {A pressure-based force and torque
  prediction technique for the study of fish-like swimming},\ }\href@noop {}
  {\bibfield  {journal} {\bibinfo  {journal} {PLoS ONE}\ }\textbf {\bibinfo
  {volume} {12}},\ \bibinfo {pages} {e0189225} (\bibinfo {year}
  {2017})}\BibitemShut {NoStop}%
\bibitem [{\citenamefont {Kheradvar}\ \emph {et~al.}(2010)\citenamefont
  {Kheradvar}, \citenamefont {Houle}, \citenamefont {Pedrizzetti},
  \citenamefont {Tonti}, \citenamefont {Belcik}, \citenamefont {Ashraf},
  \citenamefont {Linder}, \citenamefont {Gharib},\  and\ \citenamefont
  {Sahn}}]{Kheradvar_etal_2010}%
  \BibitemOpen
  \bibfield  {author} {\bibinfo {author} {\bibfnamefont {A.}~\bibnamefont
  {Kheradvar}}, \bibinfo {author} {\bibfnamefont {H.}~\bibnamefont {Houle}},
  \bibinfo {author} {\bibfnamefont {G.}~\bibnamefont {Pedrizzetti}}, \bibinfo
  {author} {\bibfnamefont {G.}~\bibnamefont {Tonti}}, \bibinfo {author}
  {\bibfnamefont {T.}~\bibnamefont {Belcik}}, \bibinfo {author} {\bibfnamefont
  {M.}~\bibnamefont {Ashraf}}, \bibinfo {author} {\bibfnamefont {J.~R.}\
  \bibnamefont {Linder}}, \bibinfo {author} {\bibfnamefont {M.}~\bibnamefont
  {Gharib}},  and\ \bibinfo {author} {\bibfnamefont {D.}~\bibnamefont {Sahn}},\
  }\bibfield  {title} {\bibinfo {title} {Echocardiographic particle image
  velocimetry: a novel technique for quantification of left ventricular blood
  vorticity pattern},\ }\href@noop {} {\bibfield  {journal} {\bibinfo
  {journal} {Journal of the American Society of Echocardiography}\ }\textbf
  {\bibinfo {volume} {23}},\ \bibinfo {pages} {86} (\bibinfo {year}
  {2010})}\BibitemShut {NoStop}%
\bibitem [{\citenamefont {Elkins}\ \emph {et~al.}(2003)\citenamefont {Elkins},
  \citenamefont {Markl}, \citenamefont {Pelc},\  and\ \citenamefont
  {Eaton}}]{Elkins_etal_2003}%
  \BibitemOpen
  \bibfield  {author} {\bibinfo {author} {\bibfnamefont {C.~J.}\ \bibnamefont
  {Elkins}}, \bibinfo {author} {\bibfnamefont {M.}~\bibnamefont {Markl}},
  \bibinfo {author} {\bibfnamefont {N.}~\bibnamefont {Pelc}},  and\ \bibinfo
  {author} {\bibfnamefont {J.~K.}\ \bibnamefont {Eaton}},\ }\bibfield  {title}
  {\bibinfo {title} {{4D} magnetic resonance velocimetry for mean velocity
  measurements in complex turbulent flows},\ }\href@noop {} {\bibfield
  {journal} {\bibinfo  {journal} {Experiments in Fluids}\ }\textbf {\bibinfo
  {volume} {34}},\ \bibinfo {pages} {494} (\bibinfo {year} {2003})}\BibitemShut
  {NoStop}%
\bibitem [{\citenamefont {Sponberg}\  and\ \citenamefont
  {Daniel}(2012)}]{SponbergDaniel2012}%
  \BibitemOpen
  \bibfield  {author} {\bibinfo {author} {\bibfnamefont {S.}~\bibnamefont
  {Sponberg}} and\ \bibinfo {author} {\bibfnamefont {T.~L.}\ \bibnamefont
  {Daniel}},\ }\bibfield  {title} {\bibinfo {title} {Abdicating power for
  control: a precision timing strategy to modulate function of flight power
  muscles},\ }\href@noop {} {\bibfield  {journal} {\bibinfo  {journal}
  {Proceedings of the Royal Society B--Biological Sciences}\ }\textbf {\bibinfo
  {volume} {279}},\ \bibinfo {pages} {3958} (\bibinfo {year}
  {2012})}\BibitemShut {NoStop}%
\bibitem [{\citenamefont {Sato}\ \emph {et~al.}(2015)\citenamefont {Sato},
  \citenamefont {Doan}, \citenamefont {Kolev}, \citenamefont {Huynh},
  \citenamefont {Zhang}, \citenamefont {Massey}, \citenamefont {van Kleef},
  \citenamefont {Ikeda}, \citenamefont {Abbeel},\  and\ \citenamefont
  {Maharbiz}}]{Sato_etal_2015}%
  \BibitemOpen
  \bibfield  {author} {\bibinfo {author} {\bibfnamefont {H.}~\bibnamefont
  {Sato}}, \bibinfo {author} {\bibfnamefont {T.~T.~V.}\ \bibnamefont {Doan}},
  \bibinfo {author} {\bibfnamefont {S.}~\bibnamefont {Kolev}}, \bibinfo
  {author} {\bibfnamefont {N.~A.}\ \bibnamefont {Huynh}}, \bibinfo {author}
  {\bibfnamefont {C.}~\bibnamefont {Zhang}}, \bibinfo {author} {\bibfnamefont
  {T.~L.}\ \bibnamefont {Massey}}, \bibinfo {author} {\bibfnamefont
  {J.}~\bibnamefont {van Kleef}}, \bibinfo {author} {\bibfnamefont
  {K.}~\bibnamefont {Ikeda}}, \bibinfo {author} {\bibfnamefont
  {P.}~\bibnamefont {Abbeel}},  and\ \bibinfo {author} {\bibfnamefont {M.~M.}\
  \bibnamefont {Maharbiz}},\ }\bibfield  {title} {\bibinfo {title} {Deciphering
  the role of a coleopteran steering muscle via free flight stimulation},\
  }\href@noop {} {\bibfield  {journal} {\bibinfo  {journal} {Current Biology}\
  }\textbf {\bibinfo {volume} {25}},\ \bibinfo {pages} {798} (\bibinfo {year}
  {2015})}\BibitemShut {NoStop}%
\bibitem [{\citenamefont {Sotiropoulos}\  and\ \citenamefont
  {Yang}(2014)}]{SotiropoulosYang_2014}%
  \BibitemOpen
  \bibfield  {author} {\bibinfo {author} {\bibfnamefont {F.}~\bibnamefont
  {Sotiropoulos}} and\ \bibinfo {author} {\bibfnamefont {X.~L.}\ \bibnamefont
  {Yang}},\ }\bibfield  {title} {\bibinfo {title} {Immersed boundary methods
  for simulating fluid-structure interaction},\ }\href@noop {} {\bibfield
  {journal} {\bibinfo  {journal} {Progress in Aerospace Sciences}\ }\textbf
  {\bibinfo {volume} {65}},\ \bibinfo {pages} {1} (\bibinfo {year}
  {2014})}\BibitemShut {NoStop}%
\bibitem [{\citenamefont {van Rees}\ \emph {et~al.}(2015)\citenamefont {van
  Rees}, \citenamefont {Gazzola},\  and\ \citenamefont
  {Koumoutsakos}}]{vanRees_etal_2015}%
  \BibitemOpen
  \bibfield  {author} {\bibinfo {author} {\bibfnamefont {W.~M.}\ \bibnamefont
  {van Rees}}, \bibinfo {author} {\bibfnamefont {M.}~\bibnamefont {Gazzola}},
  and\ \bibinfo {author} {\bibfnamefont {P.}~\bibnamefont {Koumoutsakos}},\
  }\bibfield  {title} {\bibinfo {title} {Optimal morphokinematics for
  undulatory swimmers at intermediate reynolds numbers},\ }\href@noop {}
  {\bibfield  {journal} {\bibinfo  {journal} {Journal of Fluid Mechanics}\
  }\textbf {\bibinfo {volume} {775}},\ \bibinfo {pages} {178} (\bibinfo {year}
  {2015})}\BibitemShut {NoStop}%
\end{thebibliography}%

\end{document}